\documentclass[preprint,showpacs]{revtex4}
\usepackage{epsfig}
\usepackage{graphicx}
\usepackage{bm}
\usepackage{amssymb}
\newcommand{\bmath}[1]{\mbox{\boldmath{${#1}$}}}

\newcommand{\half}{\mbox{${\textstyle \frac{1}{2}}$}}           % 1/2
\newcommand{\fmn}[2]{\mbox{${\textstyle \frac{#1}{#2}}$}}
\newcommand{\rd}{\textrm{d}}
\begin{document}
%{\flushright{\today}}
\title{Meson production in high-energy electron-nucleus scattering}
\author{G\"oran F\"aldt}\email{goran.faldt@fysast.uu.se} 
%\author{Ulla Tengblad}\email{ulla.tengblad@fysast.uu.se} 
\affiliation{ Department of physics and astronomy, 
Uppsala University,
 Box 516, S-751 20 Uppsala,Sweden }

\begin{abstract}
Experimental studies of meson production through two-photon fusion 
in inelastic electron-nucleus scattering is now under way.  A high-energy
photon radiated by the incident electron is fused with a soft
photon radiated by the nucleus. The process takes place in the
small-angle-Coulomb region of nuclear scattering. We expound the theory for 
this production process as well as its interference with 
coherent-radiative-meson production. In particular, we investigate the distortion
of the electron wave function due to multiple-Coulomb scattering.

\end{abstract}
\maketitle
%
%
%%%%%%%%%%%%%%%%%%%%%%%%%%%
%
\section{Introduction}

The PrimEx Collaboration \cite{PrimEx} aims at measuring electromagnetic 
properties of pseudoscalar mesons through the Primakoff effect, 
in 11 GeV/$c$ electron-nucleus-inelastic scattering. Experiments of this type were 
suggested some years ago by Hadjimichael and Fallieros \cite{Hadji}, and they also
elaborated their theoretical
description, within the Born approximation. 
Here, we shall develop, in a Glauber
approach \cite{RJG}, a full-fledged theory for this process. We are in particular 
concentrating on the effects of Coulomb scattering of the electron 
and on the interference of the two-photon-fusion amplitude with
the coherent-pion-photoproduction amplitude. 
The Primakoff effect on proton targets at JLab energies has been studied by 
Laget \cite{Lagetprim}.

The kinematics of the electron-nucleus-pion-production
 reaction is defined through
\begin{equation}
e^-(k_1)+ {\rm A}(p_1)\rightarrow e^-(k_2)+\pi^0(K)+{\rm} A(p_2).
\end{equation}
Our analysis is carried out for high energies and small transverse
momenta, meaning small compared with the longitudinal momenta.
In addition, the momentum transfers to pion and nucleus must be
in the Coulomb region, leading to further severe 
restrictions. In the PrimEx experiment typical energies are;
for the incident electron $E(\mathbf{k}_1)=11$ GeV, and 
for the scattered electron $E(\mathbf{k}_2)=300$ MeV. 
This implies an energy of
$\omega(\mathbf{K})=10.7$  GeV for the  pion, as for
the virtual photon initiating the pion-photoproduction process. 
The PrimEx Collaboration also plans to investigate production of
$\eta$ and $\eta'$ mesons. Our formulae are valid also in those cases
but we shall present numerical results only for pions.

The electron-nucleus production amplitude is a sum of two amplitudes;
the two-photon-fusion amplitude $ {\cal M}_{2\gamma}$, and the
coherent-photoproduction amplitude $\cal{M}_{\gamma}$.
The normalizations are chosen so that the  
 cross-section distribution takes the form
\begin{equation}
\frac{\rd \sigma}{\rd^2k_{2\bot}  \rd^2K_{\bot} \rd k_{2\|}}
  = \frac{1}{32 (2 \pi)^5 {k}_1 E(\mathbf{k}_2) K_\| M_A^2 }
    \left| {\cal M}_{2\gamma}+  {\cal M}_{\gamma}\right|^2  ,
 \label{Cross-sect-distr}
\end{equation}
 with $K_\|$ the component of pion momentum along the incident $\mathbf{k}_1$ direction.
 We neglect the recoil energy of the nucleus in comparison
 with its  mass.

The  structure of the cross-section distribution at 
small-transverse-momentum transfers is essentially determined by
the photon-exchange propagators. There are two such propagators;
one in the variable $\bmath{k}_{2\bot}$ and one in $\bmath{p}_{2\bot}$.
Each of them exhibits a Primakoff-peak structure.
%
%
%%%%%%%%%%%%%%%%%%%%%%%%%%%%%%%%%%%%%%%%%%%%%%%%%%%%%%%%%%%%%%%
%
%
\newpage
\section{Point-like-nucleus target}

\begin{figure}[ht]
%\begin{center}
\scalebox{0.90}{\includegraphics{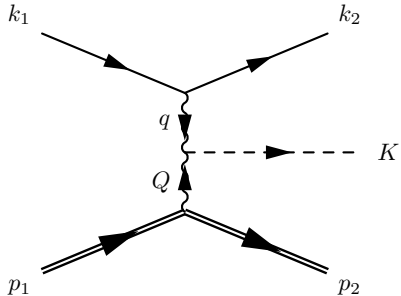} }
%\end{center}
\caption{Graph describing pion production in the Coulomb field of 
a nucleus in  inelastic-electron coherent-nucleus scattering.}\label{F1-fig}
\end{figure}

We start with the two-photon-fusion matrix element. In this initial calculation, 
the nucleus is treated as a point-like particle of zero spin. Moreover, we ignore 
the Coulomb phase of the electron wave function, which is generated by 
Coulomb  multiple scattering of the electron. In other words, only the 
Born, or tree, diagram is considered. It is displayed in Fig.~\ref{F1-fig}.

Naming the particle momenta as in Fig.~\ref{F1-fig}, the overall 
four-momentum conservation reads
\begin{equation}
	k_1+p_1=k_2+p_2 + K.
\end{equation}
The four-momenta of the intermediate photons are defined according to
\begin{eqnarray}
	k_1&=&k_2+q,\\
	p_1&=&p_2+Q,\\
	K&=&q+Q.
\end{eqnarray}
The vertex functions of the photon-fusion diagram are
\begin{eqnarray}
	{\cal M}(k_1\rightarrow k_2+q)&=&\bar{u}(k_2)ie\gamma^\mu u(k_1)\label{Vert-e}\\
	{\cal M}(p_1\rightarrow p_2+Q)&=&-iZe(p_1+p_2)^\nu,\label{Vert-Z}\\
	{\cal M}(q+Q\rightarrow K)&=& \frac{ie^2}{m_\pi}\ g_{\pi\gamma\gamma} \epsilon_{\mu\nu\rho\sigma} q^\rho Q^\sigma .
	  \label{Vert-pi}
\end{eqnarray}
The coupling constant $g_{\pi\gamma\gamma}$ of Eq.(\ref{Vert-pi}), 
with $\epsilon_{0123}=1$, is fixed 
by the anomaly of the axial-vector-current divergence. Taking its numerical value 
from the  pion-decay rate
\begin{equation}
	\Gamma(\pi^0\rightarrow \gamma \gamma)= \fmn{1}{4} \pi \alpha^2g_{\pi\gamma\gamma}^2m_\pi,
\end{equation}
gives $g_{\pi\gamma\gamma}=0.0375$.

The complete matrix element for the photon-fusion diagram of Fig.~\ref{F1-fig} becomes
 \begin{equation}
{\cal M}_{2\gamma}= ie(-iZe)(\frac{ie^2}{m_\pi} g_{\pi\gamma\gamma} )(-i)^2\
      \frac{1}{q^2} \cdot  \frac{1}{Q^2 } \
      \bar{u}(k_2)\gamma^\mu u(k_1) \epsilon_{\mu\nu\rho\sigma} q^\rho Q^\sigma (p_1+p_2)^\nu .\\
	 \label{Simple00-amp} 
\end{equation} 
This expression can be simplified. Referring to Eq.(\ref{Vert-Z}) we may write
\begin{equation}
	p_1+p_2=2p_1-Q ,
\end{equation}
where the $Q$-term contribution to Eq.(\ref{Simple00-amp}) vanishes due to gauge invariance, 
manifested by the anti-symmetry property of the $\epsilon$ factor. 
Evaluating the matrix element in the lab.~system, where  $p_1=(M_A,\mathbf{0})$,
we get
\begin{equation}
	{\cal M}_{2\gamma}= iZ\frac{e^4}{m_\pi} g_{\pi\gamma\gamma} 2M_A\
      \frac{1}{{q}^2} \cdot  \frac{1}{{Q}^2 } \
      \bar{u}(k_2)\bm{\gamma}\cdot(\mathbf{q}\times\mathbf{Q}) u(k_1)  .
	 \label{Simple11-amp} 
\end{equation}	 
So far, no approximations have been made. In the following reduction one must be careful
since the vector $\mathbf{q}=(\mathbf{q}_\bot ,q_\|)$ has a longitudinal component 
$q_\|=10.7$ GeV/$c$, and large, but a transverse component ${q}_\bot$ 
with a length rather measured in MeV/$c$.	 

Expressed in two-component form the current-matrix 	element of the three-vector 
$\bm{\gamma}=(\gamma^k)$  can be approximated as
\begin{equation}
	\bar{u}(k_2)\bm{\gamma} u(k_1)= (E_1 E_2)^{1/2} \zeta^{\dagger}_2
	 \left[ 2 \hat{k}_1 +\frac{1}{E_2}  \mathbf{k}_{2\bot} - 
	    \frac{i}{E_2}(\mathbf{k}_{2\bot}\times\bm{\sigma}) \right]\zeta_1 .
	 \label{current}	
\end{equation}	 	 
Here, we have used that the electron energies are much larger 
than the electron mass,
 $E_1,E_2\gg m_e$, and that for the scattered electron  the transverse
  momentum is much smaller than the longitudinal one, $k_{2\bot}\ll k_{2\|}$.

Now, if we take advantage of the fact that all transverse momenta are small, 
$q_{\bot},Q_{\bot}\ll q_{\|}$, that $Q_{\|}\ll q_{\|}$, and are ready to 
neglect small terms of order $E_2/E_1$, then the
two-fusion matrix element simplifies to
\begin{equation}
	{\cal M}_{2\gamma} = iZ\frac{e^4}{m_\pi}g_{\pi\gamma\gamma} 2M_A \sqrt{E_1E_2}
	\ \frac{E_1}{E_2}
	 \frac{1}{{q}^2} \cdot  \frac{1}{{Q}^2 } 
	 \cdot\zeta^{\dagger}_2
	 \left[  (  \mathbf{q}_{\bot} \times \mathbf{Q}_{\bot})\cdot \hat{k}_1 
	   -  \ \mathbf{q}_{\bot} \cdot \mathbf{Q}_{\bot}\ 
	   i\bm{\sigma}\cdot \hat{k}_1 \right]\zeta_1 .
	   	 \label{Fusion-amp} 
\end{equation}
We observe that the spin-{\sl independent} term is maximal when $ \mathbf{q}_{\bot}$ is
orthogonal to $\mathbf{Q}_{\bot}$,
whereas the spin-{\sl dependent} term is maximal when $ \mathbf{q}_{\bot}$ is 
parallel to $\mathbf{Q}_{\bot}$.

Next, we put $e^2=4\pi\alpha$ in the matrix element of Eq.(\ref{Fusion-amp}), 
square it 
and take the proper average over the electron-spin states. 
Inserting the result into Eq.(\ref{Cross-sect-distr}), gives the cross-section distribution
\begin{equation}
\frac{\rd \sigma}{\rd^2k_{2\bot}  \rd^2K_{\bot} \rd k_{2\|}}
  = \frac{1}{\pi K_\|  }
    \bigg[ \frac{Z\alpha^2}{m_\pi}g_{\pi\gamma\gamma}\ \frac{E_1}{E_2}\frac{q_{\bot}}{q^2}
    \cdot\frac{Q_{\bot}}{Q^2}\bigg]^2  ,
 \label{Cross-sect-simp}
\end{equation}
with the double-Primakoff structure clearly exhibited. Observe, that in the denominators,
$q^2$ and $Q^2$ are still squares of four-vectors. It will susequently be shown 
that in the kinematic configuration studied here this expression simplifies to
\begin{equation}
\frac{\rd \sigma}{\rd^2k_{2\bot}  \rd^2K_{\bot} \rd k_{2\|}}
  = \frac{1}{\pi K_\|  }
    \bigg[ \frac{Z\alpha^2}{m_\pi}g_{\pi\gamma\gamma}\ \frac{q_{\bot}}{\mathbf{q}_\bot^2+m_e^2}
    \cdot\frac{Q_{\bot}}{\mathbf{Q}_\bot^2+Q_\|^2}\bigg]^2  ,
\end{equation}
with $Q_\|=m_\pi^2/2K_\|$.

In the following we shall, when no confusion is possible,  use the notation 
${\cal M}_{2\gamma}$ also for the $2\times 2$ production matrix, 
of which  Eq.(\ref{Fusion-amp}) is the matrix element. 

%
%
%%%%%%%%%%%%%%%%%%%%%%%%%%%%%%%%%%%%%%%%%%%%%%%%%%%%%%%%%%%%%%%%%%%%%%%%
%
%
\newpage
\section{Impact-parameter representation}
%
%%%%%%%%%%%%%%%%%%%

We shall eventually need to introduce the Coulomb phase 
 of the high-energy electron.
This will be done in the eikonal approximation. For this task we need
the coordinate-space description of the two-photon-fusion amplitude.
In photoproduction the incident photon combines with a virtual photon
radiated by the electric field of the nucleus to form the produced pion \cite{GF-bas}.
 In the present case there are two virtual photons involved;
one radiated by the nucleus, the other radiated by the electron. Hence,
two electric fields, or potentials, are needed. We start with the 
nucleus-radiation process.

The kinematics is defined in Fig.~1. The values of the longitudinal momentum transfers
are fixed and it is important to understand their sizes. 
In the energy-conservation constraint,
\begin{equation}
	E_1(\mathbf{k}_1)+M_A=E_1(\mathbf{k}_2)+\omega(\mathbf{K})+E_A(\mathbf{Q}),
\end{equation}
we expand in powers of the longitudinal momenta. Also for the scattered electron
the longitudinal momentum is much larger than the electron mass and 
the transverse momentum, so the expansion is possible also there. 
For the nucleus, however, the recoil kinetic energy is so tiny it can
be neglected altogether. We obtain
\begin{equation}
	Q_\|=\frac{m_\pi^2}{2K_\|}.
	\label{Long-mom-tr}
\end{equation}
Since $K_\|=q_\|+Q_\|$ and $Q_\|$ extremely small we may replace $K_{\|}$
by $q_\|$ if we like. Numerically, at $K_\|=10.7$ GeV/$c$, 
the longitudinal momentum transfers for $\pi^0$- and $\eta$-meson
 production become
\begin{eqnarray}
	Q_\|(m_{\pi^0})&=& 0.85 \quad\mbox{\rm MeV/$c$}, \label{Qlong-pi}\\
	Q_\|(m_\eta)&= & 14.0 \quad\mbox{\rm MeV/$c$}.\label{Qlong-eta}
\end{eqnarray}
The value for eta production is large on a nuclear scale.

The photon radiated by the nucleus has four-momentum $Q$ and we may put
\begin{equation}
	-Q^2=\mathbf{Q}_{\bot}^2+Q_{\|}^2 ,
\end{equation}
since $Q_0$ is negligible in comparison with the space components of $Q$. 
The action of the nuclear photon
can be described by a vector field, $\mathbf{V}_A(\mathbf{r})$, which following
 Eq.~(\ref{Fusion-amp}) has the structure
\begin{equation}
	\mathbf{F}_A(\mathbf{Q}) = \frac{\mathbf{Q}_{\bot}}{\mathbf{Q}_{\bot}^2+Q_{\|}^2}= 
	  \int \rd^3r e^{-i\mathbf{Q}\cdot \mathbf{r}}\ \mathbf{V}_A(\mathbf{r}),
	  \label{F-A-qspace}
\end{equation}
with $\mathbf{Q}=\mathbf{K}-\mathbf{q}$. The index $A$ indicates that the 
field is associated with the nucleus. The potential $\mathbf{V}_A(\mathbf{r})$ 
is exactly the one that appears in real photoproduction \cite{GF-bas}. 
For a point-like-nucleus-charge distribution its analytic form is
\begin{equation}
	\mathbf{V}_A(\mathbf{r})= \frac{-1}{4\pi i}\frac{\mathbf{s}}{\left[\mathbf{s}^2+z^2\right]^{3/2}},
	\label{V-Apot}
\end{equation}
with $\mathbf{r}=(\mathbf{s},z)$.

Integration along the $z$-direction produces a vector-profile function,
\begin{equation}
	\mathbf{F}_A(\mathbf{Q}) = \frac{iq}{2\pi}
	  \int \rd^2b \;e^{-i\mathbf{Q}_\bot \cdot \mathbf{b}}\ \mathbf{\Gamma}_A(\mathbf{b},Q_{\|}),
	  \label{FA-def}
\end{equation}
with
\begin{equation}
	\mathbf{\Gamma}_A(\mathbf{b},Q_{\|}) = \frac{\hat{b}}{qb}\ \bigg[  bQ_{\|} K_1( bQ_{\|})\bigg],
	\label{Gamma-A}
\end{equation}
and $q$, with $q\approx K_\|$, the momentum of the  photon radiated by the electron. 
The function $K_1(z)$ 
is the modified Bessel function. 

The radiation by the high-energy electron is similarly described by a vector field, as 
can be seen by rewriting the propagator
\begin{eqnarray}
	q^2&=& (k_1-k_2)^2 \nonumber\\
	 &=& -\frac{k_1}{k_{2\|}}\bigg[ \mathbf{k}_{2\bot}^2+\frac{m_e^2}{k_1^2}(k_1-k_{2z})^2\bigg].
	 \label{q*qexp}
\end{eqnarray}
This structure leads to an energy-dependent electron potential. We pull out 
the energy-dependent factor $-{k_1}/k_{2z}$ and define 
\begin{equation}
	\mathbf{F}_e(\mathbf{q}) = 
	\frac{\mathbf{q}_{\bot}}{\mathbf{q}_{\bot}^2+ 
	       \displaystyle{   \frac{m_e^2}{k_1^2}\,q_{\|}^2 }  } = 
	  \int \rd^3r \;e^{i\mathbf{q}\cdot \mathbf{r}}\ \mathbf{V}_e(\mathbf{r})
	   \label{F-e-qspace}
\end{equation}
with $\mathbf{q}=\mathbf{k}_1-\mathbf{k}_2$. The index $e$ indicates that this 
vector field refers to the high-energy electron. 
  Inverting Eq.(\ref{F-e-qspace}), we get an expression
for the corresponding potential.  In analogy with 
that of  Eq.~(\ref{V-Apot}), and with $\mathbf{r}=(\mathbf{s},z)$, the electron potential reads
\begin{equation}
	\mathbf{V}_e(\mathbf{r})= \frac{1}{4\pi i}
	\frac{\mathbf{s}}{\left[\mathbf{s}^2+
	\displaystyle{\frac{k_1^2}{m_e^2}}\,z^2\right]^{3/2}}.
	\label{V-epot}
\end{equation}
The $z$-dependence enters multiplied with an enormous scale factor.

Defining the profile function through 
\begin{equation}
	\mathbf{F}_e(\mathbf{q}) = \frac{ik_1}{2\pi}
	  \int \rd^2b\; e^{i\mathbf{q}_\bot \cdot \mathbf{b}}\ \mathbf{\Gamma}_e(\mathbf{b},q_{\|}),
\end{equation}
and integrating over the $z$-variable in Eq.(\ref{F-e-qspace}) yields
\begin{equation}
	\mathbf{\Gamma}_e(\mathbf{b},q_{\|}) = \frac{-\hat{b}}{k_1b}\ 
	\bigg[  m_eb K_1( m_eb)\bigg].
	\label{Gamma-e}
\end{equation}
Due to the scaling of the $z$-variable, the longitudinal-momentum transfer 
enters in the combination $m_eq_\|/k_1\approx m_e$.

We conclude from expressions (\ref{Gamma-A}) and (\ref{Gamma-e})
for the profile functions  that the cut-off in impact parameter $b$ is $2K_\|/m_\pi^2$ in the 
nuclear-profile function and $1/m_e$ in the electron-profile function. 
For {\slshape pion production} in the PrimEx
experiment these cut-offs are numerically roughly the same, as follows 
from Eq.~(\ref{Qlong-pi}). 

At this point we draw special attention to Eq.(\ref{q*qexp}) where we extracted 
the factor $k_1/k_{2\|}=E_1/E_2$. Therefore, the Primakoff denominators, already 
encountered in the cross-section distribution of Eq.(\ref{Cross-sect-simp}),
take on a simple form when
expressed in terms of transverse momenta,
\begin{equation}
	 \frac{E_1}{E_2}\frac{1}{q^2}\cdot\frac{1}{Q^2}=\frac{1}{\mathbf{k}_{2\bot}^2+m_e^2}
	\cdot\frac{1}{\mathbf{Q}_{\bot}^2+Q_\|^2}.
\end{equation}
Remember that $Q_\|$ is fixed, and does not depend on the transverse momentum variables.

In Born approximation there are two amplitudes to be reckoned with;
\begin{eqnarray}
	G(\mathbf{q},\mathbf{Q})&=&
	  \left[ \mathbf{F}_e(\mathbf{q})\times\mathbf{F}_A(\mathbf{Q})\right]\cdot\hat{k}_1,
	   \label{Born-GqQ}\\
	 H(\mathbf{q},\mathbf{Q})&=&
	  \mathbf{F}_e(\mathbf{q})\cdot\mathbf{F}_A(\mathbf{Q}),\label{Born-HqQ}
\end{eqnarray}
in terms of which the two-photon-fusion matrix  becomes
\begin{eqnarray}
	{\cal M}_{2\gamma}&=&i {\cal N}_{2\gamma}\,
	 \bigg[ 	G(\mathbf{q},\mathbf{Q}) 
	   -   H(\mathbf{q},\mathbf{Q})\ 
	   i\bm{\sigma}\cdot \hat{k}_1 \bigg]\bigg[ 2M_A 	\sqrt{E_1E_2 }\bigg],
	   	 \label{NewFusion-amp} \\
	{\cal N}_{2\gamma}&=& Z\frac{e^4}{m_\pi}g_{\pi\gamma\gamma} .
	\label{N-factor}
\end{eqnarray}

The expressions for the amplitudes $\mathbf{F}_e(\mathbf{q})$ and $\mathbf{F}_A(\mathbf{Q})$
are given in Eqs (\ref{F-e-qspace}) and (\ref{F-A-qspace}). In momentum space
the nuclear amplitude is a direct product of amplitudes. In coordinate space,
it is a convolution of 
potentials,
\begin{equation}
	G(\mathbf{q},\mathbf{Q})=
	  \int\rd^3r_e \int\rd^3r_\pi\,  e^{i\mathbf{q}\cdot(\mathbf{r}_e -\mathbf{r}_\pi)}
	    e^{-i\mathbf{Q}\cdot\mathbf{r}_\pi}  
	     \left[\mathbf{V}_e(\mathbf{r}_e -\mathbf{r}_\pi)
	      \times \mathbf{V}_A(\mathbf{r}_\pi)\right]\cdot \hat{k}_1  , 
	      \label{BornG}
\end{equation} 
with $\mathbf{r}_e$ and $\mathbf{r}_\pi$  the coordinates of electron and
pion relative to the point nucleus. The argument of the electron potential is the
relative coordinate between electron and pion. The definition of the spin-dependent 
function $H(\mathbf{q},\mathbf{Q})$ is similarly
\begin{equation}
	H(\mathbf{q},\mathbf{Q})=
	  \int\rd^3r_e \int\rd^3r_\pi \, e^{i\mathbf{q}\cdot(\mathbf{r}_e -\mathbf{r}_\pi)}
	    e^{-i\mathbf{Q}\cdot\mathbf{r}_\pi}  
	     \left[\mathbf{V}_e(\mathbf{r}_e -\mathbf{r}_\pi)
	      \cdot \mathbf{V}_A(\mathbf{r}_\pi)\right] . 
	      \label{BornH}
\end{equation} 
%
%%%%%%%%%%%%%%%%%%%%%
%%%%%%%%%
\newpage
\section{Coulomb phase}
%
%%%%%%%%%%%%%%%%%%%%%%%
There are several improvements to the Born approximation that must be made. 
Among them we find effects of the elastic-Coulomb scattering of the electron, 
the distortion
of the pion wave within the nucleus, and the extension of the 
nuclear-charge distribution. In this section we shall treat the elastic-Coulomb
scattering of the electron.

Both initial- and final-state electrons are relativistic and move 
along the eikonal $z$-direction. The Coulomb distortion
of the electron wave function results in an eikonal-phase factor\cite{RJG},
\begin{equation}
	e^{i\chi_C(b)}=\left( \frac{2a}{b}\right)^{i\eta},
\end{equation}
with $b$ the relative impact parameter between the electron and the
point nucleus, with $a$ the cut-off radius of the Coulomb
potential, and with 
\begin{equation}
	\eta =2 Z\alpha/v  .\label{etadef}
\end{equation}
The velocity $v$ can be put to unity, and the cut-off radius
leads to an $a$-dependent phase factor common to all
amplitudes.

Changing the argument of the 
electron-profile function, replacing the electron-impact parameter
 $\mathbf{b}_e$ by $\mathbf{b}_e +\mathbf{b}_\pi$,
we get the Coulomb corrected amplitude
\begin{eqnarray}
	G(\mathbf{q},\mathbf{Q})&=& \frac{iq}{2\pi} \frac{ik_1}{2\pi}
	  \int\rd^2b_e e^{i\mathbf{q}\cdot\mathbf{b}_e}
	     \int\rd^2b_\pi 
	    e^{-i\mathbf{Q}\cdot\mathbf{b}_\pi}  
	    \Bigg[ \frac{2a}{|\mathbf{b}_e+\mathbf{b}_\pi|}\Bigg]^{i\eta}  
	          \nonumber   \\
	    && \times \left[\mathbf{\Gamma}_e(\mathbf{b}_e )
	      \times \mathbf{\Gamma}_A(\mathbf{b}_\pi)\right]\cdot \hat{k}_1  . 
	      \label{Imp-Coul-bas}  
\end{eqnarray}

The phase factor in the integrand of Eq.(\ref{Imp-Coul-bas}) couples the integrations over 
 the impact parameters $\mathbf{b}_e$ and $\mathbf{b}_\pi$,  and gives 
 rise to a nucleus form factor, with a structure that could be quite 
 complicated. It will not only depend on the absolute values of 
 the momenta $\mathbf{Q}_\bot$ and  $\mathbf{q}_\bot$, but also on their 
 scalar product. Hence, it is  expected to be significantly more complicated 
 than the Coulomb-nucleus form factor encountered in pion-nucleus 
 bremsstrahlung \cite{FT}.  The phase variation must 
 be properly investigated since Coulomb production through two-photon fusion 
 interferes with coherent
 pion-nucleus photoproduction. The Coulomb phase of the coherent pion-photoproduction
 amplitude depends only on ${q}_\bot$ and is known analytically for point-like-nuclear
 charge distributions \cite{FT}.
 
Unfortunately, in Eq.(\ref{Imp-Coul-bas})  only one angular integration can be performed analytically. The three remaining ones must be done numerically. 
We start from the basic expression, Eq.(\ref{Imp-Coul-bas}),  
\begin{eqnarray}
	G(\mathbf{q},\mathbf{Q})&=& \frac{m_eQ_\|}{(2\pi)^2} 
	  \int\rd^2b_e e^{i\mathbf{q}\cdot\mathbf{b}_e}
	     \int\rd^2b_\pi 
	    e^{-i\mathbf{Q}\cdot\mathbf{b}_\pi}  
	    \ \left(\hat{\mathbf{b}}_e \times \hat{\mathbf{b}}_\pi \right)\cdot \hat{k}_1
	          \nonumber   \\
	    &&  \cdot K_1(m_eb_e)K_1(Q_\|b_\pi)
	     \Bigg[ \frac{2a}{|\mathbf{b}_e+\mathbf{b}_\pi|}\Bigg]^{i\eta}  . 
\end{eqnarray}
Keeping the angular difference $\phi_e-\phi_\pi$ fixed, and integrating over the 
other angle gives as result
\begin{eqnarray}
	G(\mathbf{q},\mathbf{Q})&=& \frac{m_eQ_\|}{(2\pi)^2} 
	  \int_0^\infty b_e\rd b_e \ K_1(m_eb_e) \int_0^\infty b_\pi\rd b_\pi \ K_1(Q_\|b_\pi)
	  \nonumber \\ &&
	 \cdot  2\pi \int_0^{2\pi} \rd \phi [-\sin(\phi+\phi_q-\phi_Q)] 
	  \cdot J_0\left( \sqrt{(q_\bot b_e)^2+(Q_\bot b_\pi)^2
	             -2q_\bot b_eQ_\bot b_\pi \cos\phi } \right)
	\nonumber   \\&&
 \cdot	\left[ \frac{2a}{\sqrt{b_e^2+b_\pi^2+2b_eb_\pi \cos(\phi+\phi_q-\phi_Q)}}\right]^{i\eta} .
 \label{GqQ-bas}  
\end{eqnarray}
The corresponding expression for $H(\mathbf{q},\mathbf{Q})$ is obtained
making the replacement 
\begin{equation}
	-\sin(\phi+\phi_q-\phi_Q) \rightarrow \cos(\phi+\phi_q-\phi_Q).
	 \label{HqQ-bas}
\end{equation}
Remember that in $G(\mathbf{q},\mathbf{Q})$ and $H(\mathbf{q},\mathbf{Q})$
the longitudinal component of $\mathbf{q}$ is defined to be $q_\|=m_e$.

We can check Eq.(\ref{GqQ-bas}) by working out the Born approximation, i.e.~the limit of vanishing Coulomb phase,
$\eta=0$. Rewrite the sine function using the addition formula! Then, integration 
of the term proportional to  $\sin\phi$ obviously gives a vanishing contribution.
For evaluation of the cosine term we apply the identity (see Appendix A)
\begin{equation}
	2\pi \int_0^{2\pi} \rd \phi \cos(\phi)J_0\left( \sqrt{x^2+y^2-2xy\cos\phi } \right) =
		2\pi J_1(x)2\pi J_1(y),
		\label{Int01}
	\end{equation}
to get as Born approximation
\begin{eqnarray}
	G_{B}(\mathbf{q},\mathbf{Q})&=&  -\sin(\phi_q-\phi_Q) m_eQ_\| 
	  \int_0^\infty b_e\rd b_e\  J_1(q_\bot b_e)K_1(m_eb_e)
	  \int_0^\infty b_\pi\rd b_\pi \ J_1(Q_\bot b_\pi)K_1(Q_\|b_\pi)
	  \nonumber
	   \\&=& (\hat{\mathbf{q}}_\bot\times \hat{\mathbf{Q}}_\bot)\cdot\hat{k}_1\, 
	   \left[\frac{{q}_\bot}{\mathbf{q}^2}\right]  \left[\frac{{Q}_\bot}{\mathbf{Q}^2}\right] ,
\end{eqnarray}
which, as expected, agrees with Eq.(\ref{Born-GqQ}).  

Before departing on a full-blown evaluation of Eq.(\ref{Imp-Coul-bas})
we take a look at the domain of integration. Since the modified Bessel function
is exponentially damped for large values of its argument,  there is effectively
a  cut-off in $b_e$ at the inverse electron mass;
\begin{equation}
	b_{ec}=1/m_e=390 \ {\rm fm}.
	\label{be-cut}
\end{equation}
The cut-off in $b_\pi$ is at the inverse longitudinal momentum transfer
$Q_\|$. Its value depends on the meson produced. For the PrimEx experiment
at 11 GeV/$c$,
\begin{eqnarray}
 b_{\eta c}(\eta)&=&1/Q_\|(\eta)=14 \ {\rm fm};\nonumber \\
    b_{\pi c}(\pi)&=& 1/Q_\|(\pi) = 230 \ {\rm fm}.	  \label{cut-offs}
\end{eqnarray}

In the case of {\em eta production} the $b_e$-region, with cut-off from Eq.(\ref{be-cut}),
 extends much further out from the
nucleus than the $b_\eta$-region. As a consequence, it may be a good approximation 
to replace $|\mathbf{b}_e+\mathbf{b}_\eta|$ by $|\mathbf{b}_e|$, thereby factorizing
the integrand of Eq.(\ref{Imp-Coul-bas}). All integrals
can  then be  evaluated analytically, with the result
\begin{equation}
 G_{\eta}(\mathbf{q},\mathbf{Q})= 
 (\hat{\mathbf{q}}_\bot\times  \hat{\mathbf{Q}}_\bot)\cdot\hat{k}_1\, 
    \bigg[\frac{q_\bot}{\mathbf{q}^2}(aq)^{i\eta}e^{i\sigma}
     h_C(\mathbf{q})\bigg]
     \bigg[\frac{Q_\bot}{\mathbf{Q}^2}\bigg]  .
	      \label{G-0-point}  
\end{equation} 
The definition of the point-Coulomb-form factor $h_C(\mathbf{q})$, with $q_\|=m_e$, 
is given in Eq.(21) of Ref.\cite{FT}, where details of the integration
can be found. 
In the same way the expression for the spin-flip amplitude becomes,
\begin{equation}
H_{\eta}(\mathbf{q},\mathbf{Q})= 
\hat{\mathbf{q}}_\bot\cdot \hat{\mathbf{Q}}_\bot\, 
    \bigg[\frac{q_\bot}{\mathbf{q}^2}(aq)^{i\eta}e^{i\sigma}
     h_C(\mathbf{q})\bigg]
     \bigg[\frac{Q_\bot}{\mathbf{Q}^2}\bigg]  .
	      \label{H-0-point}
\end{equation} 
 $H_{\eta}(\mathbf{q},\mathbf{Q})$  differs from $G_{\eta}(\mathbf{q},\mathbf{Q})$ 
 only in the extracted kinematic factor. There is in this 
approximation, effectively, only one Coulomb amplitude, and  its $q$-dependent 
factor is also the Coulomb-form factor for meson photoproduction.

We expect Eqs (\ref{G-0-point}) and (\ref{H-0-point}) to be useful for eta 
production, but for {\em pion production}
there will be large corrections. But, since the two-photon-fusion amplitude
${\cal M}_{2\gamma}$ interferes coherently with the photoproduction
amplitude ${\cal M}_{\gamma}$ it is imperative to
 know the relative Coulomb phases of these amplitudes. 
In the photoproduction case, the virtual photon of four-momentum $q$ is
attached to a nucleon in the nucleus, and the radius of the nucleus is small 
compared with the cut-off $b_{ec}$ of Eq.(\ref{cut-offs}). The photoproduction amplitude
will hence approximately factorize, one of the factors being the
Coulomb scattering part of Eqs (\ref{G-0-point}) and (\ref{H-0-point}).

Returning now to Eq.(\ref{GqQ-bas}), we observe that 
\begin{equation}
[-\sin(\phi+\phi_q-\phi_Q)] =- \hat{\mathbf{q}}_\bot\cdot\hat{ \mathbf{Q}}_\bot\sin\phi
+	(\hat{\mathbf{q}}_\bot\times\hat{\mathbf{Q}}_\bot)\cdot\hat{k}_1\cos\phi .
\end{equation}
Integrating Eq.(\ref{GqQ-bas}) with the $\sin\phi$ factor gives a small return due to
the approximate asymmetry of the integrand during inversion $\phi\rightarrow-\phi$. The
$\cos\phi$ term, on the contrary, gives a large return due to the approximate 
symmetry of the integrand during the same  inversion. We conclude that $ G(\mathbf{q},\mathbf{Q})$ 
is most important when $\phi_q-\phi_Q=\pm \pi/2$, and vanishes identically 
when $\phi_q-\phi_Q=0, \pi$, in concordance with the properties  of 
$ G_{\eta}(\mathbf{q},\mathbf{Q})$ of Eq.(\ref{G-0-point}).
  
Next, we observe that 
\begin{equation}
[\cos(\phi+\phi_q-\phi_Q)] = \hat{\mathbf{q}}_\bot\cdot \hat{\mathbf{Q}}_\bot\cos\phi
+	(\hat{\mathbf{q}}_\bot\times\hat{\mathbf{Q}}_\bot)\cdot\hat{k}_1\sin\phi .
\end{equation}
By an argument similar to  the one of the previous paragraph we claim that 
$H(\mathbf{q},\mathbf{Q})$ is maximal when $\phi_q-\phi_Q=0, \pi$
and minimal, but not vanishing,  when $\phi_q-\phi_Q=\pm \pi/2$.

The two functions $G(\mathbf{q},\mathbf{Q})$ and $H(\mathbf{q},\mathbf{Q})$ are 
independent, but there are other ways of introducing independent form factors. 
We may, e.g., perform the $\phi$-integrations of Eqs (\ref{GqQ-bas}) and (\ref{HqQ-bas})
with factors of $\cos\phi$ and $-\sin \phi$.
In Appendix B we show that precisely these integrands appear naturally if we 
remove the vector  product (scalar for $H$) in 
$G(\mathbf{q},\mathbf{Q})$ of Eq.(\ref{Imp-Coul-bas}), with the derivative trick. 
 The new
functions that emerge  are labelled $K(\mathbf{q},\mathbf{Q})$ and $L(\mathbf{q},\mathbf{Q})$,
and in terms of them
\begin{eqnarray}
	G(\mathbf{q},\mathbf{Q}) &=& K(\mathbf{q},\mathbf{Q})\ 
	 (\hat{\mathbf{q}}_\bot\times\hat{\mathbf{Q}}_\bot)\cdot\hat{k}_1  
	 + L(\mathbf{q},\mathbf{Q})\ \hat{\mathbf{q}}_\bot\cdot\hat{ \mathbf{Q}}_\bot 
	       ,\label{Gdecomp}\\
   H(\mathbf{q},\mathbf{Q} )&=& K(\mathbf{q},\mathbf{Q})\ 
      \hat{\mathbf{q}}_\bot\cdot\hat{ \mathbf{Q}}_\bot - 
	  L(\mathbf{q},\mathbf{Q}) \ (\hat{\mathbf{q}}_\bot\times\hat{\mathbf{Q}}_\bot)\cdot\hat{k}_1 
	    \label{Hdecomp}. 
\end{eqnarray}

The integral definitions of the functions $K(\mathbf{q},\mathbf{Q})$ and 
$L(\mathbf{q},\mathbf{Q})$ are
\begin{eqnarray}
	K(\mathbf{q},\mathbf{Q})&=&  
	  \frac{m_eQ_\|}{(2\pi)^2} 
	  \int_0^\infty b_e\rd b_e \ K_1(m_eb_e) \int_0^\infty b_\pi \rd b_\pi \ K_1(Q_\|b_\pi)
	  \nonumber \\ &&
	 \cdot  2\pi \int_0^{2\pi} \rd \phi 
	  \cdot J_0\left( \sqrt{(q_\bot b_e)^2+(Q_\bot b_\pi)^2
	             -2q_\bot b_eQ_\bot b_\pi \cos\phi } \right)
	\nonumber   \\&&
 \cdot \{\cos\phi\} \cdot	\left[ \frac{2a}{\sqrt{b_e^2+b_\pi^2+2b_eb_\pi \cos(\phi+\phi_q-\phi_Q)}}\right]^{i\eta} , \label{Kffint} \\
 	L(\mathbf{q},\mathbf{Q})&=&  
	  \frac{m_eQ_\|}{(2\pi)^2} 
	  \int_0^\infty b_e\rd b_e \ K_1(m_eb_e) \int_0^\infty b_\pi \rd b_\pi \ K_1(Q_\|b_\pi)
	  \nonumber \\ &&
	 \cdot  2\pi \int_0^{2\pi} \rd \phi 
	  \cdot J_0\left( \sqrt{(q_\bot b_e)^2+(Q_\bot b_\pi)^2
	             -2q_\bot b_eQ_\bot b_\pi \cos\phi } \right)
	\nonumber   \\&&
 \cdot \{-\sin\phi\} \cdot	\left[ \frac{2a}{\sqrt{b_e^2+b_\pi^2+2b_eb_\pi \cos(\phi+\phi_q-\phi_Q)}}\right]^{i\eta} . \label{Lffint}  
\end{eqnarray}    
In the Born approximation, $\eta=0$, the $L$-function vanishes, $L_B(\mathbf{q},\mathbf{Q})=0$, 
whereas the $K$-function is reduced to
\begin{equation}
	K_{B}(\mathbf{q},\mathbf{Q})=
	   \frac{{q}_\bot}{\mathbf{q}^2}\cdot \frac{{Q}_\bot}{\mathbf{Q}^2}.\label{Kborn}	
\end{equation}

\begin{figure}[t]
%\begin{center}
\scalebox{0.45}{\includegraphics{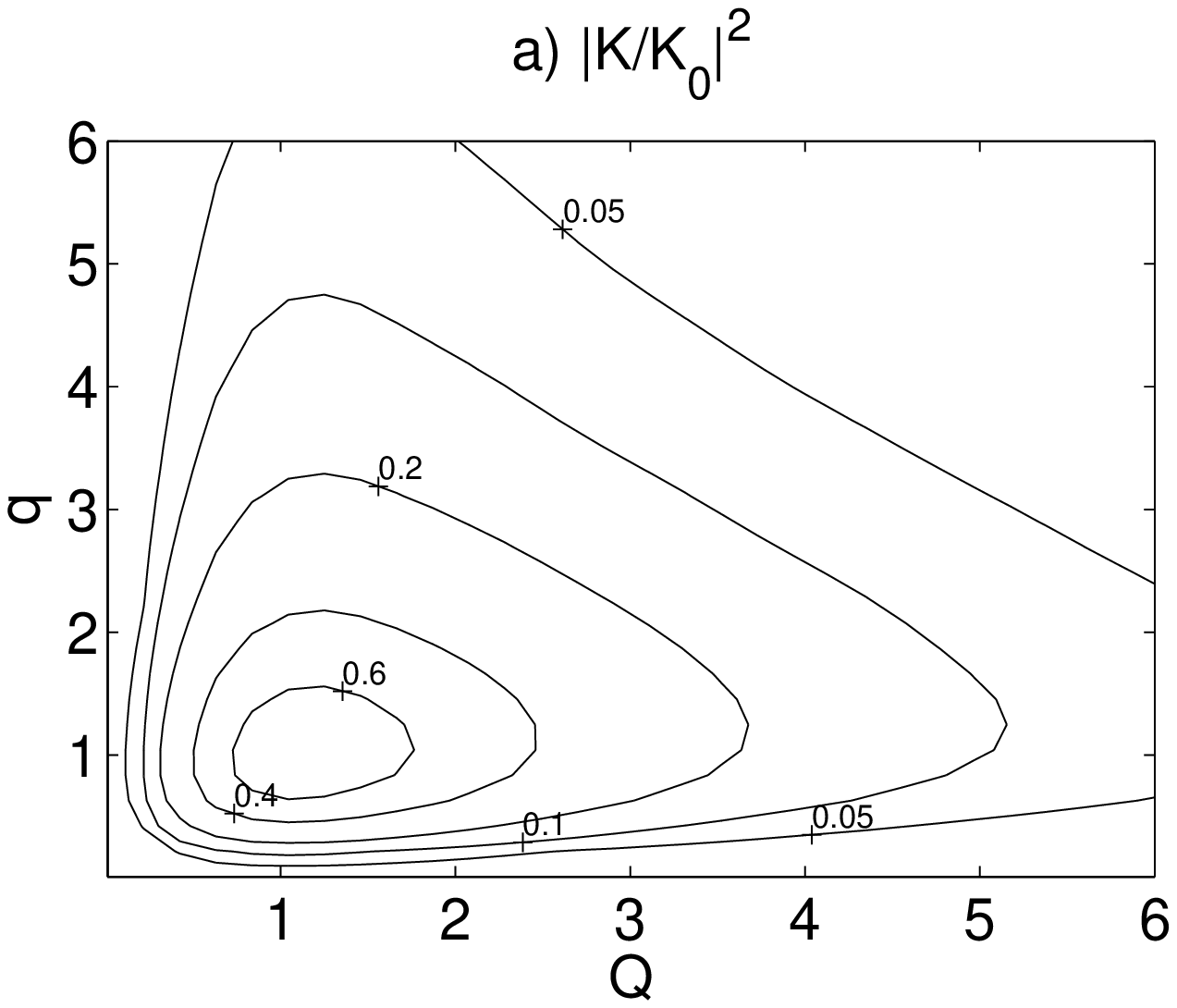} \qquad \includegraphics{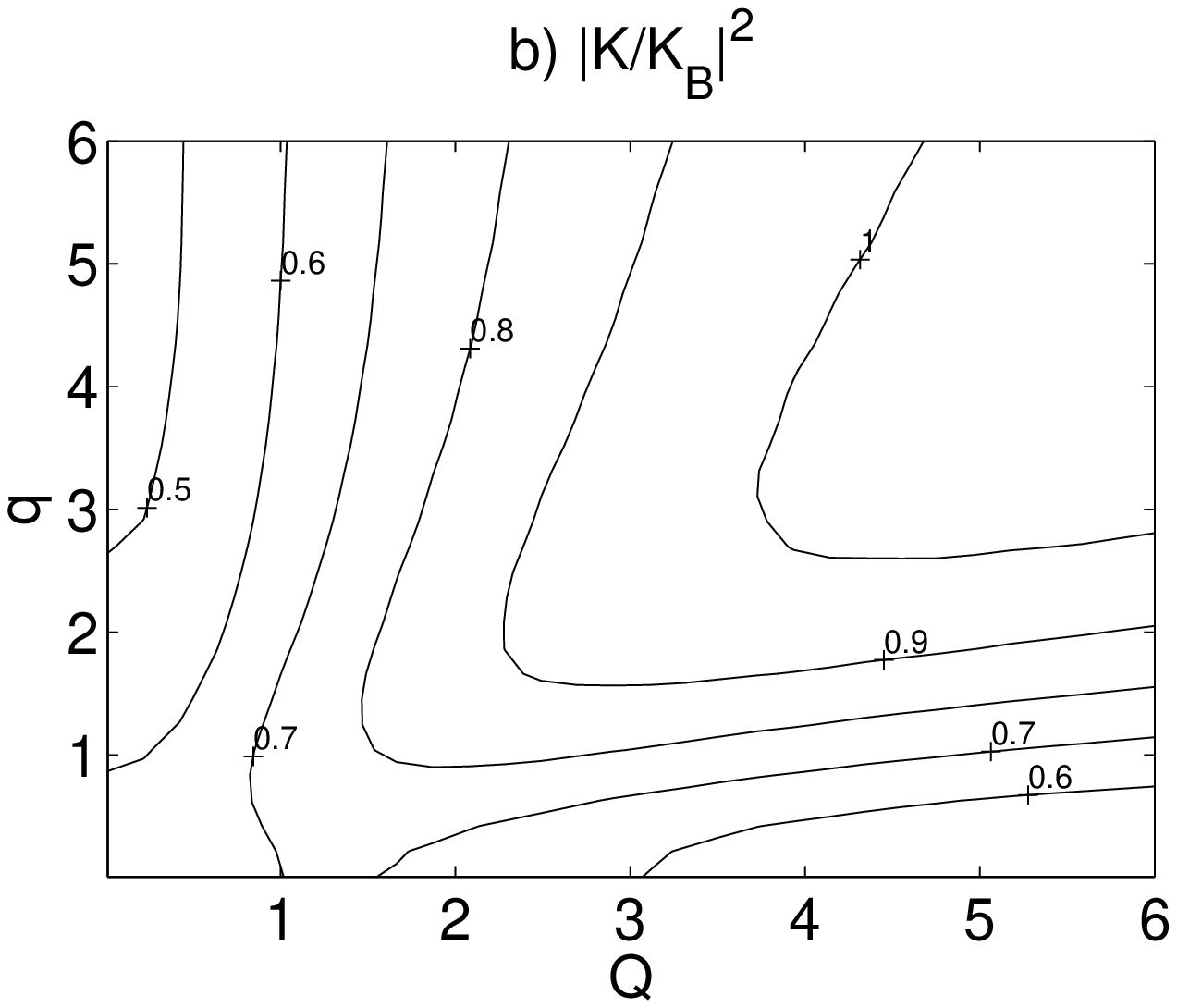}} \\
\scalebox{0.45}{\includegraphics{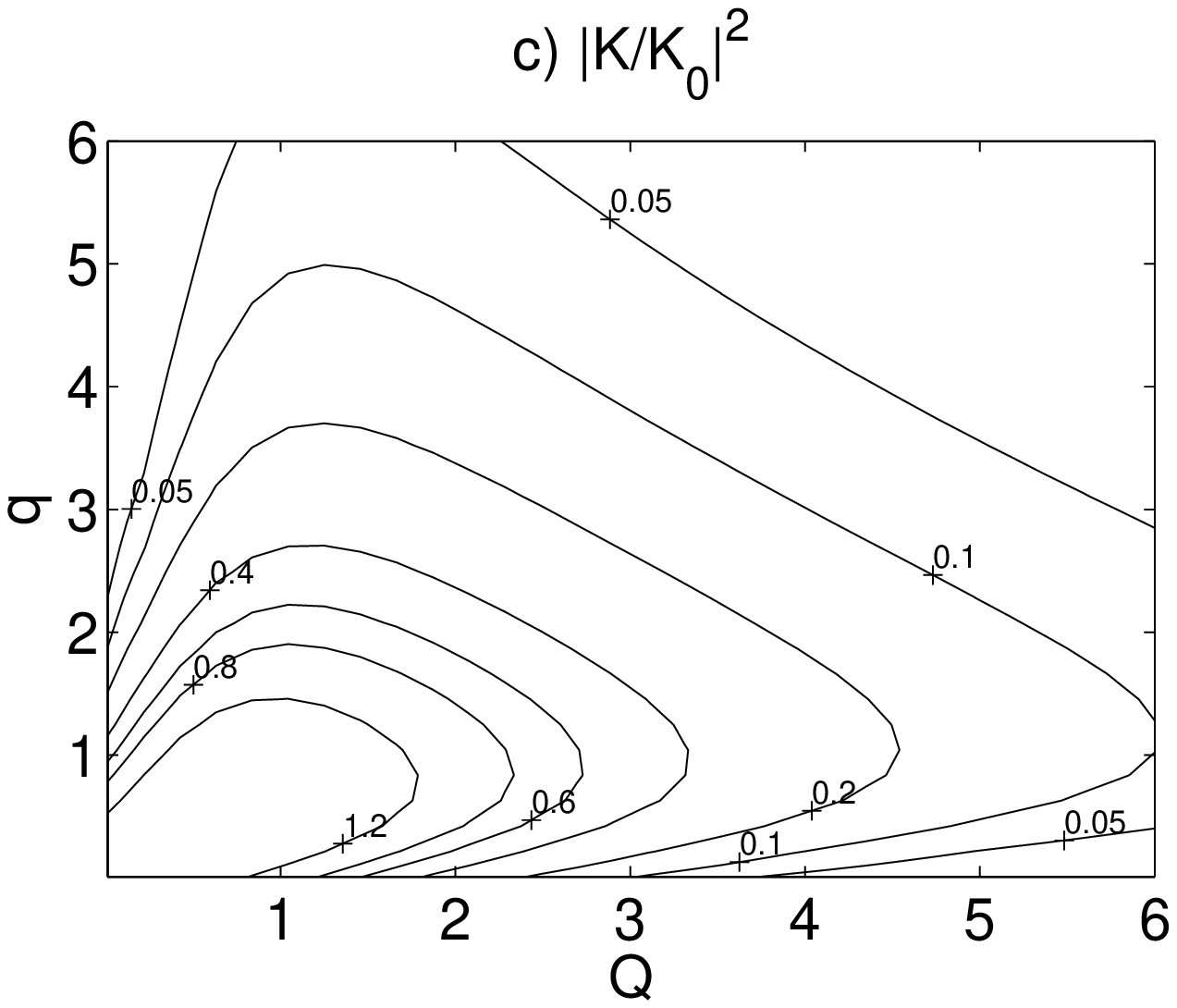} \qquad \includegraphics{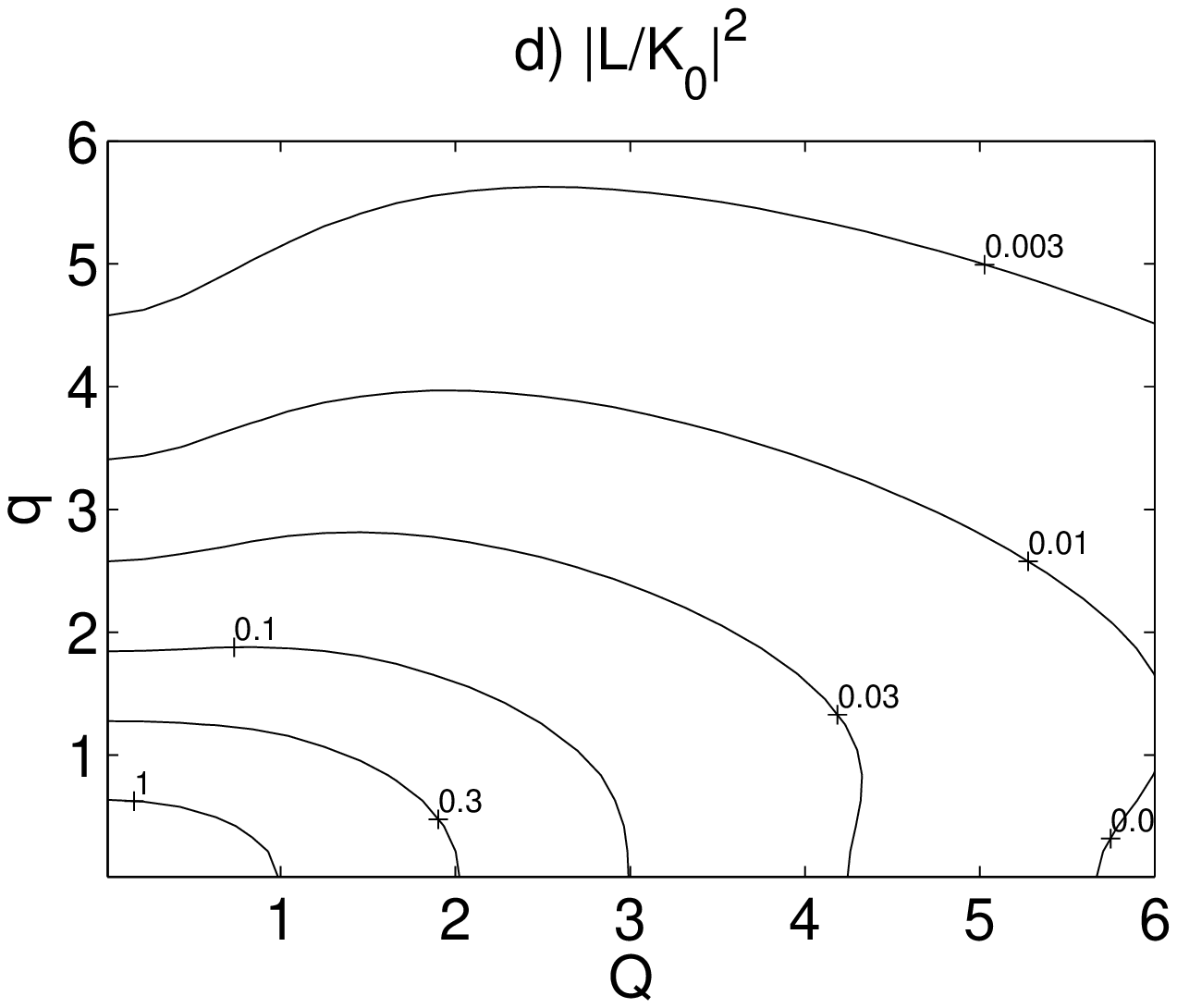}}
%\end{center}
\caption{Contour plots of the squared amplitude ratios  for  $\pi$ production on $^{208}$Pb. In panels a, c, and d the 
 ratios are with respect to $K_0=1/(4 Q_\| q_\|)$, and in panel b with respect to the 
 Born approximation $K_B(\mathbf{q},\mathbf{Q})$. In panels a, b, and d the relative $\phi$-angle 
 is $\phi_q-\phi_Q=\pi/2$,  whereas in panel c it is  $\phi_q-\phi_Q=0$. 
 The $y$-coordinate measures the momentum component $q_\bot$ in units of $q_\|$,
  and the $x$-coordinate the momentum component $Q_\bot$ in units of $Q_\|$. }
  \label{KLpi-fig}
\end{figure}

The numerical presentation of the functions $K(\mathbf{q},\mathbf{Q})$ and 
$L(\mathbf{q},\mathbf{Q})$ is difficult since 
 they depend on three variables; $q_\bot$, $Q_\bot$, and the angle 
between them $\phi_q -\phi_Q$. We are attempting this in Fig.\ref{KLpi-fig}.
In Figs \ref{KLpi-fig}a and \ref{KLpi-fig}b we present contour plots of 
the squared amplitude ratios 
for angle $\phi_q-\phi_Q=\pi/2$. In \ref{KLpi-fig}a the ratio is taken with respect 
to 
\begin{equation}
	K_0=1/(4 Q_\| q_\|),
\end{equation}
which is the maximal value of the Born amplitude (\ref{Kborn}), attained at the
double peak, and in \ref{KLpi-fig}b with respect to the Born amplitude itself.
 We conclude that in this configuration the $K$-amplitude 
vanishes, as does the Born approximation, when $Q_\bot$ or $q_\bot$ approaches 
zero. Also, the magnitude of the $K$-amplitude is considerably
smaller than that of the Born amplitude, in the double-peak region, but approaches 
the magnitude of the  Born amplitude when the values of $Q_\bot$ or $q_\bot$ becomes 
substantially larger than at the peak.

In Fig.\ref{KLpi-fig}c the relative angle is $\phi_q-\phi_Q=0$, and we notice 
a radical change. The effect of the elastic Coulomb scattering of the electron is to make
the $K$-amplitude non-vanishing when $Q_\bot$ or $q_\bot$ vanishes. In fact, 
it has a maximum at $Q_\bot=q_\bot=0$ which is stronger than the 
Born maximum at the double peak. But when we go to larger values of 
$Q_\bot$ or $q_\bot$ the amplitude values are similar to those at $\phi_q-\phi_Q=\pi/2$.

The $L$-amplitude vanishes when $\phi_q-\phi_Q=0$. Its relative magnitude
for $\phi_q-\phi_Q=\pi/2$ is graphed in Fig.\ref{KLpi-fig}d. We again see
the effect of elastic Coulomb scattering. The amplitude assumes its 
largest values when $Q_\bot$ and $q_\bot$ both take on values smaller 
than their peak values. But beyond the double peak the $L$-values are 
much smaller than the corresponding $K$-values, also for lead.

\begin{figure}[t]
\scalebox{0.45}{\includegraphics{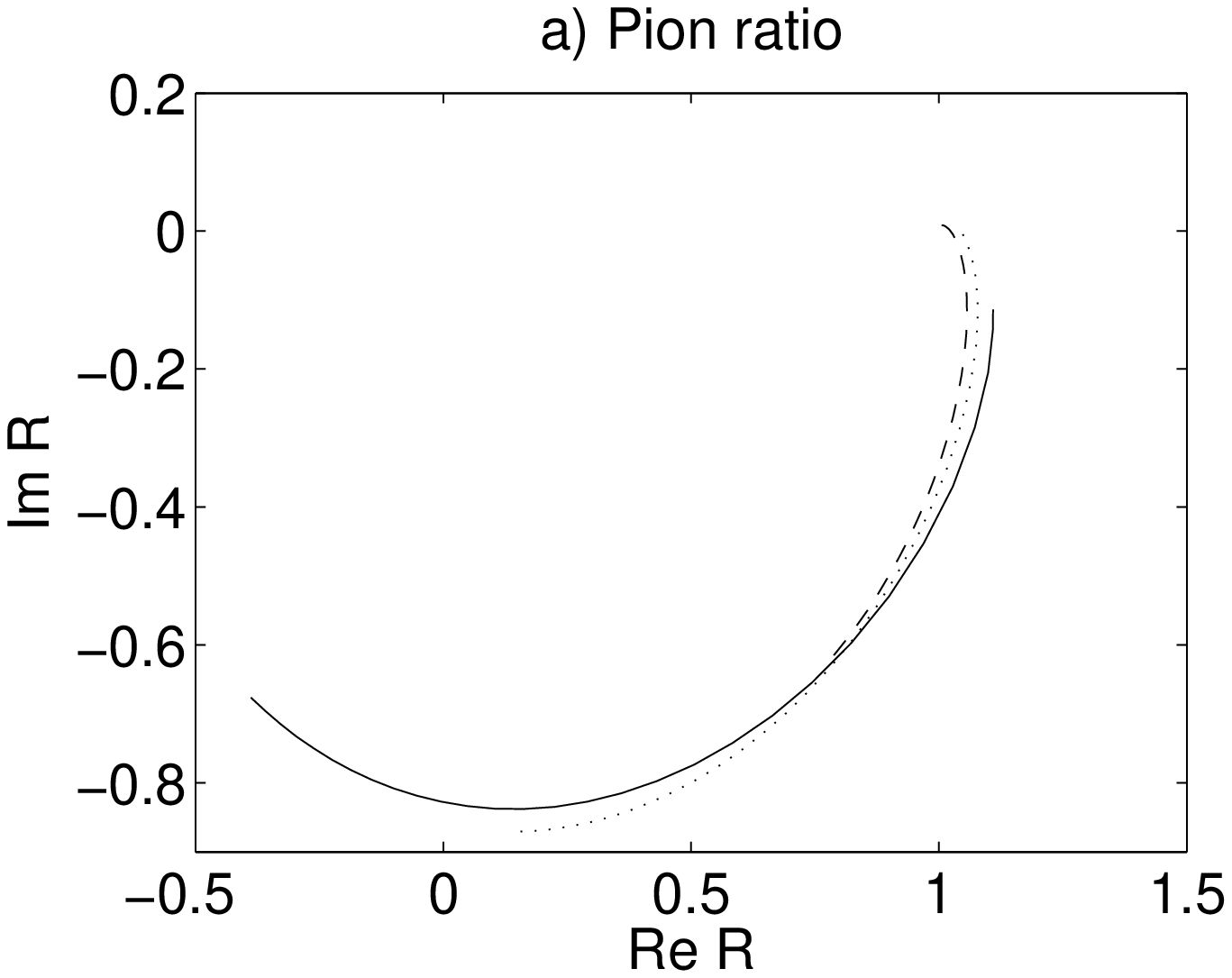} \qquad \includegraphics{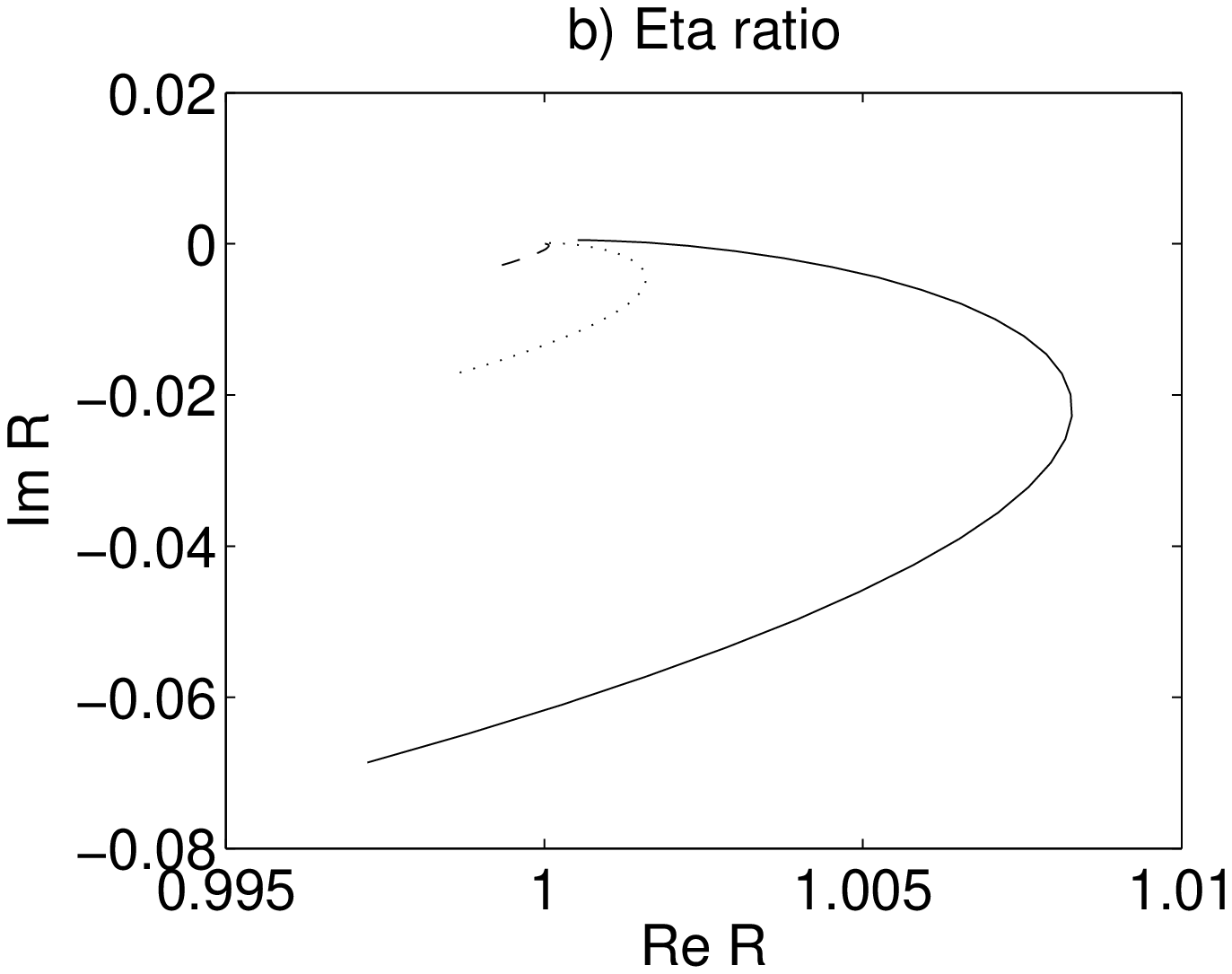}} 
\caption{  The ratio $	R(\mathbf{q},\mathbf{Q})$ defined in  Eq.(\ref{ratioeq}) 
graphed for  pion and eta production. 
Note the difference in scales! Each curve represents a fixed $Q$-value. 
The chosen values of the ratio $Q_\bot/Q_\|$ are, 
 1.0~(solid line), 2.0~(dotted), and 4.0~(dashed). The $q$-values cover the range 
$q_\bot/m_e=[0.1,10.0]$, with the largest $q$-value furthest away from unity.}
\label{Ratio-fig}
\end{figure}

Next, we return to the factorization property suggested in Eq.(\ref{G-0-point}).  
The function $L(\mathbf{q},\mathbf{Q})$ is very small for eta production,  
since the angular dependence of the phase factor in Eq.(\ref{Kffint})
may be ignored, leading to a vanishing integral. Hence, we are only interested 
in the $K$-function ratio 
\begin{equation}
	R(\mathbf{q},\mathbf{Q})=K(\mathbf{q},\mathbf{Q})\bigg/
	\bigg[\frac{q_\bot}{\mathbf{q}^2}(aq)^{i\eta}e^{i\sigma}
     h_C(\mathbf{q})\cdot
     \frac{Q_\bot}{\mathbf{Q}^2}\bigg] ,
     \label{ratioeq}
\end{equation}
with the Coulomb  form factor $h_C(\bmath{q})$ defined as \cite{FT}
\begin{equation}
	 h_C(\bmath{q})  =
    \Gamma(2-i\eta/2) \Gamma(1+i\eta/2)
	    F(i\eta/2, 1-i\eta/2 ;2; \frac{q_{\bot}^2}{q_{\bot}^2+q_{\|}^2}) ,
	      \label{def-FF-coul}
\end{equation}
and $q_{\|}=m_e$.
Furthermore, we are only interested in situations where the variable $Q_\bot$ 
is in the peak region. In Fig.\ref{Ratio-fig} we have plotted the ratio
$	R(\mathbf{q},\mathbf{Q})$
for pion production in panel 3a and for eta production in panel 3b.
The curves traced correspond to fixed  values of the ratio  $Q_\bot/Q_\|$.
The values chosen are 1.0, 2.0, and 4.0. 
Along the curves the $q$-values cover the range $q_\bot/m_e=[0.1,10.0]$, 
with the smallest $q$-value,  $q=0.1 m_e$, associated with the points nearest to unity.
We conclude that  factorization is excellent for eta production. 
For pion production  factorization is a good approximation for values of
$q$ in the peak region but becomes increasingly poor as the $q$-values increase.

%%%%%%%%%%%%%%%%%%%%%%%%%%%%%%%%%%%
%
%
%%%%%%%%%%%%%%%%%%%%%%%%%%%%%%%%%%%
\newpage
\section{Extended nuclear-charge distribution}
%
%%%%%%%%%%%%%%%%%%%%%%%%%%%%%%

The nuclear-electric field  of a point-like-nuclear-charge distribution is given in 
Eq.(\ref{V-Apot}), and controls the photon radiation by the the nucleus. In the case
of an extended nuclear-charge distribution Eq.(\ref{V-Apot}) is replaced by
\begin{equation}
	\mathbf{V}_A(\mathbf{r})= \frac{-1}{4\pi i}
	 \int \rd ^3 x'\frac{\mathbf{s}-\mathbf{s}'}{|\mathbf{r}-\mathbf{r}'|^3}\ 
	   \rho(\mathbf{x}'),
	\label{V-Apot-rho}
\end{equation} 
with charge distribution $\rho(\mathbf{x})$  normalized to unity.
If we for simplicity consider a uniform-charge distribution with radius $R_u$
then Eq.(\ref{V-Apot}) is replaced by
\begin{equation}
	\mathbf{V}_A(\mathbf{r})=	\frac{-1}{4\pi i}\frac{\mathbf{s}}{r^3}
	 \left\{
	\begin{array}{l}
	 1,\ r>R_u, \\ 
	  -2(r/R_u)^3, \ r<R_u.
	   \end{array}\right.
	\label{V-A-ext}
\end{equation} 

The profile function is implicitely defined in Eq.(\ref{FA-def}) and as a consequence 
\begin{equation}
	\mathbf{\Gamma}_A(\mathbf{b},Q_\|) = \frac{2\pi}{iq}\int_{-\infty}^\infty
	 \rd z \ e^{-iQ_\| z}\  \mathbf{V}_A(\mathbf{b},z).
\end{equation}
When $b>R_u$ the trajectory always runs outside the nucleus, and the expression for 
the point-like case applies, Eqs (\ref{Gamma-A}) and (\ref{V-A-ext}),
\begin{equation}
b>R_u:\ \ \	
\mathbf{\Gamma}_A(\mathbf{b},Q_{\|}) = \frac{\hat{b}}{qb}\ \bigg[  bQ_{\|} K_1( bQ_{\|})\bigg]
	\label{Gamma-A>}.
\end{equation}
The case $b<R_u$ is more complicated and integrals cannot be 
done analytically;
\begin{eqnarray}
b<R_u:\ \ \	
\mathbf{\Gamma}_A(\mathbf{b},Q_{\|})& =& \frac{\hat{b}}{qb}\ \bigg[  -\frac{2}{Q_\|R_u}
 \frac{b^2}{R_u^2} \sin(Q_\|\sqrt{R_u^2-b^2} ) +
   bQ_{\|} K_1( bQ_{\|})  \nonumber \\
  && -b\sqrt{1-b^2/R_u^2} -b^2\int_0^{\sqrt{R_u^2-b^2}} \rd z\frac{\cos(Q_\|z)-1}{(b^2+z^2)^{3/2}} \bigg]
	\label{Gamma-A<}.
\end{eqnarray}
Here, the first term inside the brackets is the contribution from the $z$-integration
over the uniform interior.
The second term is the total contribution for a point-like charge distribution, and
the last two terms removes from it the contribution to the $z$-integration over 
the nuclear interior with a point-like charge distribution.

In the production of eta mesons we expect a substantial part to come from the nuclear
interior, and hence, a proper description of the nuclear-charge distribution 
is certainly called for. For production of pi mesons, on the other hand, very little
comes from the nuclear interior and it should be sufficient to treat the
nuclear-charge distribution as point like.  Since, we are mostly interested in the 
pion case we do not present any numerical results for extended-charge distributions.

The profile function of the electron is not affected by this discussion.

%%%%%%%%%%%%%%%%%%%%%%%%%%%%%%%%%%%
%
%
%%%%%%%%%%%%%%%%%%%%%%%%%%%%%%%%%%%
\newpage
\section{Pion multiple scattering }
%
%%%%%%%%%%%%%%%%%%%%%%%%%%%%%%%%%%%%%%
The pion is produced at $\mathbf{r}=(\mathbf{b},z)$ and moves from there with constant 
$\mathbf{b}$ to $z=\infty$. Along this trajectory it can scatter
off the nucleons in the nucleus. Following Glauber  \cite{RJG}
this scattering introduces a pion-wave-function distortion 
represented by the factor  $P(\mathbf{r})$, which for 
heavy nuclei takes the form
\begin{equation}
	P(\mathbf{r})=\exp[ -\frac{\sigma '}{2}T(\mathbf{b},z)], 
\end{equation}
with  the target-thickness function
\begin{equation}
	T(\mathbf{b},z)=\int_z^\infty \rd z'n(\mathbf{b},z'),
\end{equation}
and $n(\mathbf{r})$ the nucleon density in the nucleus. This density is normalized 
to A, the number of nucleons in the nucleus. For the structure of the density,
we refer to Eq.(\ref{Dens-def}) of Appendix C.

The integration along the path of the pion gives  for the uniform nucleon-density
distribution,
\begin{equation}
	T(\mathbf{b},z)=
 \left\{ \begin{array}{ll}
  0 ;\ \ & b>R_u\\
      2n_0\sqrt{R_u^2-b^2}; \ \ & b<R_u,\ z<-\sqrt{R_u^2-b^2}\\
     n_0\sqrt{R_u^2-b^2}-z;\ \ & b<R_u,\ |z|<\sqrt{R_u^2-b^2}\\
     0;\ \ & b<R_u,\ z>\sqrt{R_u^2-b^2}
 \end{array} \right. \ ,
\end{equation}
with $n_0$ the nucleon density.

The pion rescattering is taken into account by replacing the nucleus
potential of Eq.(\ref{V-Apot}) by
\begin{equation}
	\mathbf{V}_A(\mathbf{r})= \frac{-1}{4\pi i}\frac{\mathbf{b}}{|\mathbf{b}^2+z^2|^{3/2}}
	\exp[ -\frac{\sigma '}{2}T(\mathbf{b},z)].
	\label{V-Apot-pi}
\end{equation}
with $\mathbf{r}=(\mathbf{b},z)$. This extension could have some influence on
eta production, since a large fraction of this production takes place inside
or near the nucleus. For pion production it should be negligible.
%%%%%%%%%%%%%%%%%%%%%%%%%%%%%%%%%%
%
%
%%%%%%%%%%%%%%%%%%%%%%%%%%%%%%%%%%%
\newpage
\section{Coherent-pion photoproduction}
%
%%%%%%%%%%%%%%%%%%%%%%%%%%%%%%%%%%%

\begin{figure}[ht]
%\begin{center}
\scalebox{0.90}{\includegraphics{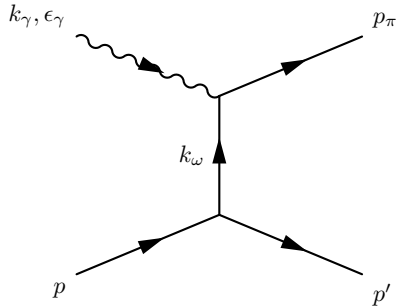} }
%\end{center}
\caption{Graph describing pion-photoproduction on the nucleon through $t$-cannel omega exchange.}
\label{Omegafig}
\end{figure}

The neutral pion can also be produced in a coherent-nuclear-photoproduction process,
a contribution which interferes coherently with the two-gamma-fusion process.
Since we are interested in a coherent-nuclear-reaction amplitude its main contribution
is generated by the spin-isospin non-flip term
of the underlying nucleon amplitude.
This nucleon amplitude can be modelled by the $\omega$-exchange 
diagram of Fig.~\ref{Omegafig}. In writing down the contribution of
this diagram we are particularly anxious 
 to get the correct Lorentz structure. 

The  $\omega$-decay vertex into $\pi^0\gamma$ in the graph of Fig.~\ref{Omegafig}
is described by the matrix element
\begin{equation}
	{\cal M}(\omega\rightarrow \pi^0\gamma)=ie\frac{g_{\omega\pi\gamma}}{m_\pi}
	   \varepsilon_{\mu\nu\rho\sigma}
	   \epsilon_\gamma^\mu \epsilon_\omega^\nu k_\gamma^\rho k_\omega^\sigma \ ,
\end{equation}
where we may replace $k_\omega$ by $p_\pi$. 
The decay width of $\Gamma(\omega\rightarrow\pi^0\gamma)=760$ keV sets
the coupling strength to $g_{\omega\pi\gamma}=0.322$. 

The vetex for $\omega$ coupling to a nucleon takes the form
\begin{equation}
		{\cal M}(\omega NN)=-ig_{\omega NN}\bar{u}(p')\left[\not\!{\epsilon}_\omega
		+\frac{\kappa_\omega}{4M}
		( \not\!{\epsilon}_\omega\not\!{k}_\omega - \not\!{k}_\omega  \not\!{\epsilon}_\omega)
		\right]u(p).
		\label{OmegNNdef}
\end{equation}
Numerical values for the coupling constants are $g_{\omega NN}=15.9$ and 
$\kappa_\omega=0$. Furthermore, in coherent-nuclear production the nucleon spin-flip
amplitudes are heavily supressed. Therefore, if we rewrite Eq.(\ref{OmegNNdef}) 
with the aid of the Gordon decomposition 
\begin{equation}
		{\cal M}(\omega NN)=-ig_{\omega NN}
		\bar{u}(p')\bigg[ \frac{1}{2M}(p+p')\cdot{\epsilon}_\omega  +
		\frac{1}{4M}\left(  \not\!{\epsilon}_\omega \not\!{k}_\omega - \not\!{k}_\omega  
		 \not\!{\epsilon}_\omega\right) \bigg]u(p),
		 \label{OmegNNredef}
\end{equation}
it follows that only the current part survives in the nuclear matrix element. The {\em effective} 
nucleon-photoproduction amplitude responsible for the corresponding coherent-nuclear amplitude becomes
\begin{equation}
		{\cal M}(\gamma N\rightarrow \pi^0 N)=-ie \frac{g_{\omega\pi\gamma}}{m_\pi}
		g_{\omega NN}\frac{1}{2M}\bar{u}(p')u(p)  \frac{1}{t-m_{\omega}^2}
		\varepsilon_{\mu\nu\rho\sigma}
	  k_\gamma^\mu \epsilon_\gamma^\nu k_\omega^\rho (p+p')^\sigma. 
	  \label{OmeNUC}
\end{equation}
The matrix element for $\pi^0$ photoproduction is  equal  for protons and neutrons.

In order to make this expression useful at higher energies we  
replace the omega-pole factor
\begin{equation}
	P_\omega= \frac{1}{t-m_{\omega}^2}
\end{equation}
by its Reggeized version \cite{Laget1}, i.e.,
\begin{equation}
	R_\omega(s,t)= \left(\frac{s}{s_0}\right)^{\alpha_\omega(t)-1}
	\frac{\pi\alpha_\omega'}{\sin(\pi\alpha_\omega(t))}\cdot\frac{1}{\Gamma(\alpha_\omega(t))}
	\cdot\frac{S_\omega +e^{-i\pi\alpha_\omega(t)}}{2}.
\end{equation}
The signature of the omega trajectory is positive, $S_\omega=1$, and its parametrization has
 been determined from 
photoproduction  experiments by Guidal et al.\cite{Laget1} to be
\begin{equation}
	\alpha_\omega(t)= 0.44 +0.9 t,
\end{equation}
with  $t$ in units of (GeV/$c$)$^2$ and so that $\alpha_\omega'=0.9$ (GeV/$c$)$^2$. 
In this study a strong emphasis is on reproducing the  minimum in the 
cross-section distribution at $-t=0.5$ (GeV/$c)^2$, whereas in our nuclear application
the relevant $t$-values are substantially smaller. 

Specializing to the lab.~system where the initial proton is at rest, we get
as effective nucleon-photoproduction matrix
\begin{equation}
		{\cal M}^L(\gamma N\rightarrow \pi^0 N)=i e\frac{g_{\omega\pi\gamma}}{m_\pi}
		g_{\omega NN} R_\omega(s,t)\, 
		(\mathbf{k}_\gamma \times \bm{\epsilon}_\gamma )\cdot\mathbf{k}_{\pi}
	\left[ 2M\right] .
	  \label{LabNUC}
\end{equation}
As we can see there is no spin-dependent contribution.
For insertion into the coherent-nuclear-photoproduction amplitude only the 
forward-angle-pion photoproduction amplitude is needed, so
we can set $t=0$. Furthermore, we have chosen to present numerical 
results for photon energies of $k_\pi=10.7$ GeV/$c$, corresponding to 
$s=21$ (GeV/$c$)$^2$. At this energy, and at $t=0$ (GeV/$c$)$^2$,
\begin{equation}
	L_\omega(s,t)= \frac{g_{\omega\pi\gamma}}{m_\pi}g_{\omega NN}R_\omega(s,t) 
	= 5.86-i 4.85 \quad \mbox{\rm (GeV/$c$)$^{-3}$} .
\end{equation}

Also  rho exchange contributes to neutral-pion photoproduction, but
with amplitudes of opposite signs for proton and neutron.
Taking the parameters of the rho trajectory again from Ref.\cite{Laget1},
and a rho-decay width of $\Gamma(\rho\rightarrow\pi^0\gamma)=68$ keV,
we get at $s=21$ (GeV/$c$)$^2$ and $t=0$ (GeV/$c$)$^2$,
\begin{equation}
	L_\rho^p(s,t)=-	L_\rho^n(s,t)= \frac{g_{\rho\pi\gamma}}{m_\pi}g_{\rho NN}R_\rho(s,t)
	= 0.169-i 0.198 \quad \mbox{\rm (GeV/$c$)$^{-3}$} .
\end{equation}
The rho parameters are much smaller than the omega parameters, in part because
the coupling constant $g_{\rho NN}=3.25$ is much smaller than the
corresponding omega-coupling constant. In addition when we calculate the nuclear amplitude
the omega contributions from protons and neutrons add, yielding a multiplicative factor $Z+N=A$, 
whereas the rho contributions subtract, yielding a smaller
multiplicative factor $Z-N$. Consequently, it is legitimate to ignore the rho contribution
altogether, and we shall do so.
\begin{figure}[ht]
%\begin{center}
\scalebox{0.90}{\includegraphics{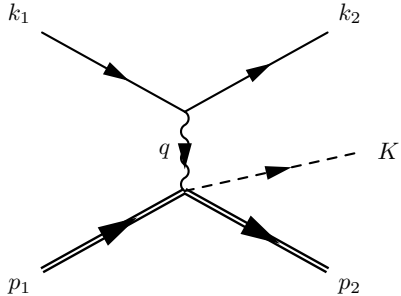} }
%\end{center}
\caption{Graph describing coherent-pion photoproduction in the Coulomb field of 
a nucleus in  inelastic-electron coherent-nucleus scattering.}\label{F2-fig}
\end{figure}

The nuclear-photoproduction diagram is shown in Fig.~\ref{F2-fig}. The external kinematics 
is the same as in Fig.~\ref{F1-fig}, and the electron vertex the same as in 
Eq.(\ref{Vert-e}). The photon propagator is also the same, but for
the photoproduction vertex we need the nuclear version of Eq.(\ref{OmeNUC}). 
However, the similarity with two-photon fusion does not end here.
The coupling of the omega to the nucleon, Eq.(\ref{OmegNNdef}), is the same as the coupling
of the photon to the proton, Eq.(\ref{Vert-e}), 
provided we neglect the magnetic terms, which we do. 	In Born approximation the
photoproduction amplitude has thus the same structure as the two-photon-fusion
amplitude, Eq.(\ref{Simple11-amp}). Before Reggeization, the photoproduction
amplitude is obtained by changing the photon-nucleus coupling constant $Ze$ 
into the omega-nucleus 
coupling constant $Ag_{\omega NN}$, and the photon propagator $1/Q^2$ into the omega
propagator $1/(Q^2-m_\omega^2)$.

But there is also 
a nuclear form factor. From Eq.(\ref{Long-mom-tr}) we know that the momentum
transfer to the nucleus along the direction $\hat{k}_1$ of the incident 
electron is fixed
\begin{equation}
  	Q_\|=(\mathbf{K}-\mathbf{q})\cdot\hat{k}_1=\frac{m_\pi^2}{2K_\|},
\end{equation}
whereas the component $\mathbf{Q}_\perp$ orthogonal to this direction varies.
The nuclear form factor must also reflect  the distortion of the  
pion wave due to hadronic scattering within the nucleus. Moreover, the nucleon amplitude
of Eq.(\ref{LabNUC}) vanishes in the forward direction, as it proportional
to $(\mathbf{q}\times \bm{\epsilon}_\gamma )\cdot\mathbf{Q}$.
The mathematical form of the appropriate nuclear form
factor $F_s(\mathbf{Q})$ is given in Ref.\cite{GF-bas}, and its operational
definition is through
\begin{equation}
	\mathbf{h}\cdot\mathbf{Q}_\perp F_s(\mathbf{Q})=
	- i\int\rd^3x e^{-i\mathbf{Q}\cdot\mathbf{x}} 
	\exp[-\half \sigma_\pi'\int_z^\infty \rd z'n(\mathbf{b},z')]\ 
	\mathbf{h}\cdot\bm{\nabla}_b n(\mathbf{b},z),
	\label{Hadron-form}
\end{equation}
where $\mathbf{h}=\mathbf{q}\times \bm{\epsilon}_\gamma $
 is a vector in the impact plane. The pion is produced at the point
$(\mathbf{b},z)$ in the nucleus where the nucleon density is $n(\mathbf{b},z)$. 
The incoming photon wave is undistorted whereas the distortion of the outgoing pion 
wave is determined by the pion-nucleon total cross section 
$\sigma_\pi'=\sigma_\pi(1-i\alpha_\pi)$.

To explain this result we start from the omega-exchange factor
\begin{equation}
	\frac{\mathbf{h}\cdot\mathbf{Q}}{\mathbf{Q}^2+m_\omega^2}=
	\int\rd^3r e^{-i\mathbf{Q}\cdot\mathbf{r}}
	\left(-i\mathbf{h}\cdot\bm{\nabla}\frac{1}{4\pi r}e^{-m_\omega r}\right).
\end{equation}
Then we fold with the nucleon density $n(\mathbf{r})$ and take the limit
of large omega mass,
\begin{eqnarray}
	& &\int\rd^3r e^{-i\mathbf{Q}\cdot\mathbf{r}}\int\rd^3r_N
	\left(-i\mathbf{h}\cdot\bm{\nabla}_\mathbf{r}\frac{1}{4\pi |\mathbf{r}-\mathbf{r}_N|}
	e^{-m_\omega |\mathbf{r}-\mathbf{r}_N|}\right)
	n(\mathbf{r}_N)\nonumber \\
	&=&\int\rd^3r e^{-i\mathbf{Q}\cdot\mathbf{r}}\int\rd^3r_N
	\frac{1}{4\pi
	|\mathbf{r}-\mathbf{r}_N|} e^{-m_\omega |\mathbf{r}-\mathbf{r}_N|}
	\left(-i\mathbf{h}\cdot\bm{\nabla}_{r_N} n(\mathbf{r}_N)\right)\nonumber \\
	&=&\frac{1}{m_\omega^2}\int\rd^3r e^{-i\mathbf{Q}\cdot\mathbf{r}}
	\left(-i\mathbf{h}\cdot\bm{\nabla}_{b} n(\mathbf{r})\right).
\end{eqnarray}
Adding  distortion due to pion scattering leads to Eq.(\ref{Hadron-form}). 
In this derivation we have used the standard expression for the omega pole 
instead of its Reggeized version. This should not change too much since we 
essentially only need the forward value of the nuclear form factor.
 
All this adds up to a nuclear photoproduction matrix element, which in 
Born approximation is given by
\begin{eqnarray}
	{\cal M}_\gamma(e^- A\rightarrow e^- \pi^0 A)&=&
	{\cal M}^\mu(e^-\rightarrow \gamma e^-)
	 \frac{-ig_{\mu\nu}}{q^2}{\cal M}^\nu(\gamma A\rightarrow  \pi^0 A)\nonumber\\
	 &=&-i(ie)(ie)L_\omega(s_{\gamma})F_s(\mathbf{Q})
	 \frac{1}{q^2}\ \bar{u}(k_2)\gamma^\nu u(k_1)
	 \varepsilon_{\mu\nu\rho \sigma } q^\mu K^\rho (p_1+p_2)^\sigma,
\end{eqnarray}
where $s_\gamma$ is the energy of the underlying nucleon photoproduction
process. In our application $s_\gamma=21$ (GeV/$c$)$^2$.  
  Evaluating the matrix element 
in the lab.~system gives a formula similar to the one for 
the two-gamma matrix element, Eq.(\ref{Simple11-amp}), 
\begin{equation}
	{\cal M}_\gamma = ie^2 2M_A L_\omega(s_{\gamma})F_s(\mathbf{Q})
	\  \frac{1}{q^2}\ 
	\bar{u}(k_2)\bm{\gamma}\cdot(\mathbf{q}\times \mathbf{Q})u(k_1),
	 \label{Simple1g}
\end{equation}
where we have replaced $\mathbf{K}$ by $\mathbf{Q}$.

We have previously shown that in the impact parameter plane the integration
over the radiation potential extends out to 396 fm. The extension in impact
parameter space  of the  photoproduction amplitude is limited to the nuclear region,
which never extends more than 7 fm from the origin. In the folding of the two 
potentials we may neglect the extension of the nucleus and factorize the two amplitudes.
Thus, to a good approximation the electron multiple scattering is contained 
in the expression
\begin{equation}
	{\cal M}_{\gamma} = i{\cal N}_{\gamma}\, M(\mathbf{q},\mathbf{Q})  
	     \bigg[(\hat{\mathbf{q}}_{\bot} \times \hat{\mathbf{Q}}_{\bot})\cdot \hat{k}_1 
	     -\hat{\mathbf{q}}_{\bot} \cdot \hat{\mathbf{Q}}_{\bot}
	     \, i\bm{\sigma}\cdot \hat{k}_1\bigg] \bigg[ 2M_A\sqrt{E_1E_2}\bigg],
	   	 \label{Photo-pr-II-amp} 
\end{equation}
where ${\cal N}_{\gamma}$ represents the product of coupling constants
\begin{equation}
	{\cal N}_{\gamma}=e^2\, \frac{g_{\omega\pi\gamma}}{m_\pi}
	g_{\omega NN}R_\omega(s,0)m_\omega^2F_s(\mathbf{Q}),
	\label{N_const_g}
\end{equation}
and
\begin{equation}
	M(\mathbf{q},\mathbf{Q}) =\frac{Q_\bot}{m_\omega^2}\cdot
	\frac{q_\bot(aq)^{i\eta}e^{i\sigma}}{\mathbf{q}^2}\, h_C(\mathbf{q}).
	\label{M_func}
\end{equation}
This amplitude  includes  multiple-pion scattering through the nuclear 
form factor $F_s(\mathbf{Q})$. In the present application we may evaluate
$F_s(\mathbf{Q})$ at $\mathbf{Q}=0$. For a uniform nuclear-density distribution
an analytic expression for $F_s(0)$ can be obtained. This is demonstrated 
in Appendix C. Also, we should keep in mind that in the form factor $h_C(\mathbf{q})$ 
the argument is interpreted as 
$\mathbf{q}=(\mathbf{q}_\bot, m_e)$.

We have defined the function $	M(\mathbf{q},\mathbf{Q})$ of Eq.(\ref{M_func}) 
so that its origin in omega exchange becomes obvious, since 
$m_\omega^2-Q^2\approx m_\omega^2$.  
The dimensionless factor $R_\omega(s,0)m_\omega^2$ then determines
 the strength of the omega-exchange 
pole  as obtained from the Regge-pole model, and
\begin{equation}
	R_\omega(s,0)m_\omega^2=0.095-i0.078.
\end{equation}
 
%
%
%%%%%%%%%%%%%%%%%%%%%%%%%%%%%%%%%%%
\newpage
\section{Cross-section distributions}
%
%%%%%%%%%%%%%%%%%%%%%%%%%%%%%%%%%%%%%%

The cross section distribution is given by Eq.(\ref{Cross-sect-distr}) of Sect.1,
\begin{equation}
\frac{\rd \sigma}{\rd^2k_{2\bot}  \rd^2K_{\bot} \rd k_{2\|}}
  = \frac{1}{32 (2 \pi)^5 {k}_1 E(\mathbf{k}_2) K_\| M_A^2 }
    \left| {\cal M}_{2\gamma}+  {\cal M}_{\gamma}\right|^2  ,
 \label{Cross-sect-distr-2}
\end{equation}
with $K_\|$ the component of pion momentum along the incident $\mathbf{k}_1$ direction.
The expression for the two-photon fusion  matrix 
  is found in Eqs (\ref{NewFusion-amp})
 and (\ref{N-factor})
 \begin{eqnarray}
	{\cal M}_{2\gamma}&=& -i {\cal N}_{2\gamma}
	 \left[ 	G(\mathbf{q},\mathbf{Q}) 
	   -   H(\mathbf{q},\mathbf{Q})\ 
	   i\bm{\sigma}\cdot \hat{k}_1 \right]\bigg[2M_A \sqrt{E_1 E_2}\bigg],
	  	 \label{NewFusion-amp-2} \\
	{\cal N}_{2\gamma}&=& Z\frac{e^4}{m_\pi}g_{\pi\gamma\gamma} .
	\label{N-factor-2}
\end{eqnarray}
The functions $G(\mathbf{q},\mathbf{Q})$ and $H(\mathbf{q},\mathbf{Q})$ 
are defined in Eqs (\ref{GqQ-bas}) and  (\ref{HqQ-bas}),  but we shall rather
employ their decompositions  into  functions $K(\mathbf{q},\mathbf{Q})$ 
and $L(\mathbf{q},\mathbf{Q})$ of Eqs (\ref{Gdecomp}) and (\ref{Hdecomp}).

The expression for the photoproduction matrix is found in Eqs (\ref{Photo-pr-II-amp}) 
and (\ref{N_const_g})
\begin{eqnarray}
	{\cal M}_{\gamma} &=&  i{\cal N}_\gamma	\,  M(\mathbf{q},\mathbf{Q}) 
	 \left[  ( \hat{\mathbf{q}}_{\bot} \times \hat{\mathbf{Q}}_{\bot})\cdot \hat{k}_1 
	   -   \hat{\mathbf{q}}_{\bot} \cdot \hat{\mathbf{Q}}_{\bot}\, 
	   i\bm{\sigma}\cdot \hat{k}_1 \right]  \bigg[2M_A \sqrt{E_1 E_2}\bigg],	 
	       \label{Photo-fin} \\ 
	{\cal N}_{\gamma} &=& e^2\, \frac{g_{\omega\pi\gamma}}{m_\pi}
	g_{\omega NN}R_\omega(s,0)m_\omega^2F_s(\mathbf{0}),	   
\end{eqnarray}
with the function $M(\mathbf{q},\mathbf{Q})$ as defined in Eq.(\ref{M_func}). 
In the peak region the form factor $F_s(\mathbf{Q})$ can be evaluated at 
zero momentum transfer, 
where it gives the effective number of nucleons in omega exchange.
According to Appendix C we have for lead, with pion rescattering included, 
$F_s(\mathbf{0})=0.70A$.

From this exposition we derive for the unpolarized cross section the expression
\begin{equation}
\frac{\rd \sigma}{\rd^2k_{2\bot}  \rd^2K_{\bot} \rd k_{2\|}}
  = \frac{1}{\pi  K_\|  }\left( \frac{Z \alpha^2g_{\pi\gamma\gamma}}{m_\pi}\right)^2
    \bigg[\, \bigg|  K(\mathbf{q},\mathbf{Q}) - 
       {\cal N}_R M(\mathbf{q},\mathbf{Q})\bigg|^2  
       +\bigg|  L(\mathbf{q},\mathbf{Q})\bigg|^2 \, \bigg],
 \label{Cross-sect-distr-3}
\end{equation}
with ${\cal N}_R$ the ratio of coupling constants of the photoproduction and the
 two-photon fusion processes,
\begin{equation}
	{\cal N}_R =\frac{g_{\omega\pi\gamma}g_{\omega NN}}{e^2g_{\pi\gamma\gamma}}
	\, R_\omega m_\omega^2\,F_s(\mathbf{0})/Z. 
\end{equation}
The expression for ${\cal N}_R$ is obvious, considering the similarities
of the two processes, provided we remember that $R_\omega m_\omega^2$ is the strength
of the omega pole, according to the Regge modell.
An interesting feature of Eq.(\ref{Cross-sect-distr-3}) is that
the photoproduction $M$-amplitude only interferes with the  
two-photon-fusion $K$-amplitude.  
  
First, we would like to determine the relative size of the photoproduction contribution.
We do this in the Born approximation where
\begin{eqnarray}
	K_B(\mathbf{q},\mathbf{Q})  &=&  
	   \frac{q_\bot}{\mathbf{q}^2} \cdot\frac{Q_\bot}{ \mathbf{Q}^2} ,
   \label{Born-KqQ-2}     \\
	  L_B(\mathbf{q},\mathbf{Q})&=& 0,   \label{Born-LqQ-2}\\
 M_B(\mathbf{q},\mathbf{Q})&=& 
  \frac{q_\bot}{\mathbf{q}^2} \cdot \frac{Q_\bot}{ m_\omega^2} . \label{Born-MqQ-2}  
\end{eqnarray}
In this approximation the $q$-dependence is the same for the two-photon-fusion 
and the photoproduction amplitudes. The ratio between the coupling-constants factors is
\begin{equation}
	\left| {\cal N}_R\right|=3.08\times 10^{-3}.
\end{equation}
For large values of $Q$ the photoproduction contribution dominates. Only in 
the vicinity of the peak can the two-photon fusion contribution be measured. 
At the peak, where $Q_\bot=Q_\|=0.85$ MeV/$c$, the enhancement factor becomes
\begin{equation}
	\frac{m_\omega^2}{2Q_\|^2}=4.22 \times 10^5,
\end{equation}
making the two-photon-fusion amplitude 1300 times stronger than the photoproduction
amplitude. Consequently, at the peak a clean measurement of the pion-decay 
constant should be possible. 
This conclusion is not affected by our ignorance of the relative sign
between ${\cal N}_{2\gamma}$ and ${\cal N}_{\gamma}$. 

Even if the two-photon-fusion contribution is dominant in the vicinity of the peak, 
the theoretical analysis is non-trivial. There are two contributions represented
by the $K$- and $L$-amplitudes. They both depend on the relative angle between 
$\mathbf{Q}_\bot$ and $\mathbf{q}_\bot$. Only the $K$-amplitude interferes with the
photoproduction amplitude. The $K$-amplitude is dominant at all angles but is 
considerably smaller than its Born approximation. 
The $L$-amplitude vanishes in the Born approximation, but is for lead at some
angles as important as the $K$-amplitude  and at other angles unimportant.
So the theoretical description of pion production in the peak region is certainly
non-trivial.  A closer scrutiny of  Fig.2 emphasizes this picture.

If we aim for a determination of the pion decay constant $g_{\pi\gamma\gamma}$,  
it is mandatory that the transverse momentum component of the pion, $Q_\bot$, 
be in the peak region, $Q_\bot\approx Q_\|$, but 
the transverse momentum of the electron, $q_\bot$, need not be in its peak region, 
$q_\bot\approx m_e$. The point, however, is that if we are in this peak region as well 
the cross section is at its strongest.

%
%
%
%%%%%%%%%%%%%%%%%%%%%%%%%%%%%%%%%%%
\newpage
\begin{acknowledgements}
It is my great fortune that I could rely on the help of Bengt R. Karlsson 
in my fight with MATLAB. I would also like to thank Colin Wilkin for useful comments.
\end{acknowledgements}
%
%
%%%%%%%%%%%%%%%%%%%%%%%%%%%%%%%%%%%
\newpage
\section{Appendix A}
The integral encountered in Eq.(\ref{Int01}) is a special case of an integral by 
Sonine and Gegenbauer \cite{Watson}. It is most easily established by expanding the 
Bessel function using Neumann's (or Gegenbauer's) addition theorem. 
Since the integral is not always listed in standard tables of integrals, e.g.\ missing in
Gradshteyn and Ryzhik \cite{GRint}, we shall here show how to evaluate it. The method 
is simple and employs the same technique that is used for other 
impact-parameter integrations of Sect.~IV.

Define the function $F(\mathbf{q},\mathbf{Q})$ as
\begin{equation}
	F(\mathbf{q},\mathbf{Q})=\int_0^{2\pi}\rd\varphi_a \int_0^{2\pi}\rd\varphi_b
	\ (\hat{\mathbf{a}}\times\hat{\mathbf{b}})\cdot\hat{\mathbf{h}}
	\exp[i\mathbf{q}\cdot\mathbf{a}- i\mathbf{Q}\cdot\mathbf{b}].
	\label{Def-angle-int}
\end{equation}
Here, the vectors $\mathbf{q}$, $\mathbf{Q}$, $\mathbf{a}$, and $\mathbf{b}$, are all in the 
$xy$-impact plane and $\mathbf{h}$ along the positive $z$-direction, i.e.\ along the 
normal to the impact plane. 
Integrations are over the angles $\varphi_a$ and $\varphi_b$. The steps of 
integration are as follows
\begin{eqnarray}
	F(\mathbf{q},\mathbf{Q})&=&\frac{1}{ab} 
	(\bm{\nabla}_q \times \bm{\nabla}_Q)\cdot\hat{\mathbf{h}}
	\int_0^{2\pi}\rd\varphi_a \int_0^{2\pi}\rd\varphi_b
	\exp[i\mathbf{q}\cdot\mathbf{a}- i\mathbf{Q}\cdot\mathbf{b}] \nonumber \\
	&=& \frac{1}{ab} 
	   (\bm{\nabla}_q \times \bm{\nabla}_Q)\cdot\hat{\mathbf{h}}\ 
	     2\pi J_0(qa) 2\pi J_0(Qb) \nonumber \\
	 &=& 2\pi J_1(qa) 2\pi J_1(Qb)\ 
	    (\hat{\mathbf{q}}\times\hat{\mathbf{Q}})\cdot\hat{\mathbf{h}}
	      \nonumber\\ 
	  &=&\sin(\varphi_Q-\varphi_q)\  2\pi J_1(qa) 2\pi J_1(Qb).    \label{Newintres}
\end{eqnarray}

The second more complicated approach is that of Sect.~3.We keep the angle
$\varphi=\varphi_a-\varphi_b$ fixed while integrating over $\varphi_b$. 
During the integration over  $\varphi_b$, the factor
\begin{equation}
	(\hat{\mathbf{a}}\times\hat{\mathbf{b}})\cdot\hat{\mathbf{h}}=-\sin\varphi
\end{equation}
stays constant as well as the scalar product 
$\hat{\mathbf{a}}\cdot\hat{\mathbf{b}}=\cos\varphi$. The exponential can
be manipulated into
\begin{equation}
	E=\mathbf{q}\cdot\mathbf{a}- \mathbf{Q}\cdot\mathbf{b}
	  =\sqrt{A}\cos(\varphi_A+\varphi_b),
\end{equation}
with
\begin{equation}
	A=(qa)^2+(Qb)^2-2qaQb\cos(\varphi_q-\varphi_Q-\varphi),
\end{equation}
and the angle $\varphi_A$ defined by
\begin{equation}
	\tan \varphi_A=-\frac{qa\sin(\varphi_q-\varphi)-Qb\sin(\varphi_Q)}
	{qa\cos(\varphi_q-\varphi)-Qb\cos(\varphi_Q)},
\end{equation}
a knowledge which is not really needed. After integration over $\varphi_b$ we get
\begin{equation}
	F(\mathbf{q},\mathbf{Q})=2\pi\int_0^{2\pi}\rd\varphi [-\sin(\varphi+\varphi_q-\varphi_Q)]
	 J_0(\sqrt{A(\varphi)}),
\end{equation}
where now 
\begin{equation}
	A(\varphi)=(qa)^2+(Qb)^2-2qaQb\cos\varphi. 
\end{equation}
	Furthermore, from the angle-addition theorem  
\begin{equation}
	\sin(\varphi+\varphi_q-\varphi_Q)=\sin(\varphi)\cos(\varphi_q-\varphi_Q)
	+\cos(\varphi)\sin(\varphi_q-\varphi_Q),
\end{equation}
we may conclude that the term proportional to $\sin\varphi$ vanishes on integration.
Comparing with expression (\ref{Newintres}) it follows that
\begin{equation}
	2\pi \int_0^{2\pi} \rd \phi \cos(\phi)J_0\left( \sqrt{x^2+y^2-2xy\cos\phi } \right) =
		2\pi J_1(x)2\pi J_1(y),
		\label{Bess-int}	
\end{equation}
as promised. If we need integrals of this type but 
with higher powers of $\cos\phi$ in the integrand we 
start from Eq.(\ref{Def-angle-int}) with integrands containing higher powers
of $(\hat{\mathbf{a}}\times\hat{\mathbf{b}})\cdot\hat{\mathbf{h}}$.

%%%%%%%%%%%%%%%%%%%%%%%%%%%%%%%%%%%
\newpage
\section{Appendix B}
The alternative approach to the nuclear form factor of Eq.(\ref{Imp-Coul-bas}) 
is to replace  $\bm{b}_e$ and $\bm{b}_\pi$ by derivatives
\begin{eqnarray}
	\bm{b}_e \rightarrow -i\bm{\nabla}_q & = & -i\left[ \hat{q}\frac{\partial}{\partial q}
	+(\hat{k}\times \hat{q})\frac{1}{q}\frac{\partial}{\partial \varphi_q} \right],  \\
	\bm{b}_\pi\rightarrow +i\bm{\nabla}_Q & = & +i\left[ \hat{Q}\frac{\partial}{\partial Q}
	+(\hat{k}\times \hat{Q})\frac{1}{Q}\frac{\partial}{\partial \varphi_Q} \right].
\end{eqnarray}
Here, $\bm{q}$ and $\bm{Q}$ are two-dimensional vectors in the impact parameter 
plane, and $\hat{k}$ is along the normal to this plane, i.e.~along the positive 
$z$-direction. As a consequence the polar-angular basis vectors are
\begin{eqnarray}
	\bm{e}_{\varphi_q} &=& \hat{k}\times \hat{q}\\
	\bm{e}_{\varphi_Q} &=& \hat{k}\times \hat{Q}.
\end{eqnarray}
In the evaluation of Eq.(\ref{Imp-Coul-bas}) and its counterpart we 
meet the differential operators
\begin{eqnarray}
	{\mathcal D}_K  &=& \frac{\partial}{\partial q}\frac{\partial}{\partial Q}
	  + \frac{1}{qQ}\frac{\partial}{\partial \varphi_q}\frac{\partial}{\partial \varphi_Q}  ,   \\
	{\mathcal D}_L  &=& \frac{1}{q}\frac{\partial}{\partial \varphi_q}\frac{\partial}{\partial Q}
	  -\frac{1}{Q}\frac{\partial}{\partial \varphi_Q}\frac{\partial}{\partial q}    , 
\end{eqnarray}
in terms of which 
\begin{eqnarray}
		\bm{b}_e \cdot 	\bm{b}_\pi & \rightarrow & \bm{\nabla}_q \cdot \bm{\nabla}_Q =
		   \hat{q}\cdot \hat{Q}\ {\mathcal D}_K 
		   +\hat{k}\cdot(  \hat{q}\times \hat{Q})\  {\mathcal D}_L ,\\
		 \hat{k}\cdot(\bm{b}_e \times 	\bm{b}_\pi )
		         & \rightarrow & \hat{k}\cdot(\bm{\nabla}_q \times\bm{\nabla}_Q)=
		         \hat{k}\cdot(  \hat{q}\times \hat{Q})\  {\mathcal D}_K
		         - \hat{q}\cdot \hat{Q}\ {\mathcal D}_L   .
\end{eqnarray}
 
 The basic functions $G(\mathbf{q},\mathbf{Q})$ and $H(\mathbf{q},\mathbf{Q})$ 
 are replaced by two other functions $K(\mathbf{q},\mathbf{Q})$ and 
 $L(\mathbf{q},\mathbf{Q})$, where $L(\mathbf{q},\mathbf{Q})$ vanishes when the 
 elastic Coulomb is turned off. 
 
Going back to Eq.(\ref{GqQ-bas}), once more,  we define  the new form factor 
$K(\mathbf{q},\mathbf{Q})$ as
\begin{eqnarray}
	K(\mathbf{q},\mathbf{Q})&=& -{\mathcal D}_K \ \Bigg\{ 
	  \frac{m_eQ_\|}{(2\pi)^2} 
	  \int_0^\infty \rd b_e \ K_1(m_eb_e) \int_0^\infty \rd b_\pi \ K_1(Q_\|b_\pi)
	  \nonumber \\ &&
	 \cdot  2\pi \int_0^{2\pi} \rd \phi 
	  \cdot J_0\left( \sqrt{(q b_e)^2+(Q b_\pi)^2
	             -2q b_eQ b_\pi \cos\phi } \right)
	\nonumber   \\&&
 \cdot	\left[ \frac{2a}{\sqrt{b_e^2+b_\pi^2+2b_eb_\pi \cos(\phi+\phi_q-\phi_Q)}}\right]^{i\eta}
 \Bigg\} .
 \label{Kff}  
\end{eqnarray} 
For the form factor $L(\mathbf{q},\mathbf{Q})$ we substitute the operator 
${\mathcal D}_L$ for ${\mathcal D}_K$. Finally, 
remember that in this appendix $q$ and $Q$ represent the transverse momentum components.

Pulling out a differential operator from inside an integral is a useful approach
when the remaining integral can be done by hand. That is not the case here but
the technique nevertheless gives new insights. We start with $L(\mathbf{q},\mathbf{Q})$.
In differentiating the elastic-Coulomb-phase factor with respect to an angle 
we note that 
%
% $\partial/\partial \varphi_q=-\partial/\partial \varphi_Q=\partial/\partial \varphi$.
%
$\partial_{\varphi_q}=-\partial_{\varphi_Q}=\partial_{\varphi}$.
Then we make a partial integration and let $\partial_\varphi$ act on the Bessel function.
The operator ${\mathcal D}_L$ now acts on the Bessel function, and with
\begin{equation}
	X=(q b_e)^2+(Q b_\pi)^2 -2q b_eQ b_\pi \cos\phi
\end{equation}
we get
\begin{equation}
	{\mathcal D}_L J_0(\sqrt{X})= \frac{-1}{qQ}\partial_\varphi(q\partial_q+Q\partial_Q)
	J_0(\sqrt{X})=b_eb_\pi \sin\phi\ J_0(\sqrt{X}).
\end{equation}
The differentiation of the Bessel function involves applying the operator 
$q\partial_q + Q\partial_Q$ to a homogeneous second order polynomial, $X$,
an operation resulting in a factor of two.

Repeating these operations for $K(\mathbf{q},\mathbf{Q})$ we get 
\begin{equation}
	{\mathcal D}_K J_0(\sqrt{X})= \left(\partial_q\partial_Q-\frac{1}{qQ}\ \partial_\varphi^2\right)
	J_0(\sqrt{X})=-b_eb_\pi \cos\phi\ J_0(\sqrt{X}).
\end{equation} 

After reintroducing our earlier notation for the transverse components, 
the integral definitions of the functions $K(\mathbf{q},\mathbf{Q})$ and 
$L(\mathbf{q},\mathbf{Q})$ become as spelled out 
in Eqs (\ref{Kffint}) and (\ref{Lffint}).

%
%
%%%%%%%%%%%%%%%%%%%%%%%%%%%%%%%%%%%
\newpage
\section{Appendix C}
The pion-hadronic form factor $F_s(\mathbf{Q})$ of Eq.(\ref{Hadron-form})
can be expressed as
\begin{equation} F_s(\mathbf{Q})= -2\pi  \int_{-\infty}^{\infty}\rd z e^{-iQ_\| z}
 \int_0 ^\infty b\rd b \ \frac{J_1(Q_\perp b)}{Q_\perp b}
 \exp\bigg[-\half \sigma_\pi'\int_z^\infty \rd z'n(b,z')\bigg]
 \ b\frac{\partial n(b,z)}{\partial b}.
	\label{Hadron-form-explicite}
\end{equation}
In the Coulomb region, the variation with momentum transfer of this
form factor is  weak. Therefore, it is sufficient to consider 
the forward direction, setting $\mathbf{Q}=0$. For a uniform nuclear density,
the nucleon density is
\begin{equation}
	n(\mathbf{r})=n_0\theta(R_u-r),\label{Dens-def}
\end{equation}
with $R_u=r_0 A^{1/3}$ the uniform nuclear radius, and the central density
\begin{equation}
	n_0=\frac{3}{4\pi r_0^3}=\frac{3A}{4\pi R_u^3}.
\end{equation}
If we observe that
\begin{equation}
	\frac{\partial n(r)}{\partial b}= -n_0 \frac{b}{r} \delta(r-R_u),
\end{equation}
the integrations in Eq.(\ref{Hadron-form-explicite}) could easily be done
by hand. After a few  straightforward manipulations we get
\begin{equation}
	F_s(0)=A\bigg[ 1+\frac{3}{2\zeta}\left(1- \frac{2}{\zeta^2} + \frac{2}{\zeta^2}(1+\zeta)
	 e^{-\zeta }\right) \bigg],
\end{equation}
with 
\begin{equation}
	\zeta=\sigma_\pi' n_0 R_u.
\end{equation}
A typical value of the pion-nucleon total cross section is $\sigma_\pi=26$ mb.
For $\alpha_\pi=0$ the numerical value of the form factor  
becomes, for lead,  $F_s(0)=0.70A$. In the absence of pion scattering, $\zeta=0$, 
and  $F_s(0)=A$.

%%%%%%%%%%%%%%%%%%%%%%%%%%%%%%%%%%

\end{document}